\documentclass[twocolumn]{aastex62}

\usepackage[utf8]{inputenc}
\usepackage{amsmath}

\newcommand{\kai}{\ion{K}{1}}%
\newcommand{\ali}{\ion{Al}{1}}%
\newcommand{\nai}{\ion{Na}{1}}%
\newcommand{\cai}{\ion{Ca}{1}}%
\newcommand{\naline}{\ion{Na}{1}{} 1.14 \micron}
\newcommand{\newnaline}{\ion{Na}{1}{} 1.27 \micron}
\newcommand{\pabeta}{Pa~$\beta$}

\newcommand{\alphak}{$\alpha_{K}$}
\newcommand{\msun}{M$_{\sun}$}
\newcommand{\kms}{km s$^{-1}$}%
\newcommand{\plmn}{$\pm$}%
\newcommand{\imfi}{$x_{1}$}%
\newcommand{\imfii}{$x_{2}$}%

\received{2019 January 15}
\accepted{2019 March 19}
\submitjournal{ApJ}

\begin{document}
\title{Initial Mass Function Variation in two Elliptical Galaxies using Near-Infrared Tracers}
\author{R. Elliot Meyer}
\affiliation{Department of Astronomy \& Astrophysics, University of Toronto, 50 St. George Street, Toronto, ON, M5S 3H4, Canada}
\author{Suresh Sivanandam}
\affiliation{Department of Astronomy \& Astrophysics, University of Toronto, 50 St. George Street, Toronto, ON, M5S 3H4, Canada}
\affiliation{Dunlap Institute for Astronomy \& Astrophysics, University of Toronto, 50 St. George Street, Toronto, ON, M5S 3H4, Canada}
\author{Dae-Sik Moon}
\affiliation{Department of Astronomy \& Astrophysics, University of Toronto, 50 St. George Street, Toronto, ON, M5S 3H4, Canada}

\correspondingauthor{R. Elliot Meyer}
\email{meyer@astro.utoronto.ca}

\begin{abstract}
Using integral field spectroscopy, we demonstrate that gravity-sensitive absorption features in the zJ-band (0.9--1.35 \micron) can constrain the low-mass stellar initial mass function (IMF) in the cores of two elliptical galaxies, M85 and M87. Compared to the visible bands, the near-infrared (NIR) is more sensitive to light from low-mass dwarf stars, whose relative importance is the primary subject of the debate over IMF variations in nearby galaxies. Our analysis compares the observed spectra to the latest stellar population synthesis models by employing two different methods: equivalent widths and spectral fitting. We find that the IMF slopes in M85 are similar to the canonical Milky Way IMF with a median IMF-mismatch parameter $\alpha_{K} = 1.26$. In contrast, we find that the IMF in M87 is steeper than a Salpeter IMF with $\alpha_{K} = 2.77$. The derived stellar population parameters, including the IMF slopes, are consistent with those from recent results in the visible bands based on spectroscopic and kinematic techniques. Certain elemental abundances, e.g. Na and Fe, have dramatic effects on the IMF-sensitive features and therefore the derived IMF slopes. We show evidence for a high [Na/H] $\sim$ 0.65 dex in the core of M85 from two independent \nai{} absorption features. The high Na abundance may be the result of a recent galactic merger involving M85. This suggests that including [Na/H] in the stellar population model parameters is critical for constraining the IMF slopes in M85. These results confirm the viability of using NIR absorption features to investigate IMF variation in nearby galaxies. 
\end{abstract}


\section{INTRODUCTION}\label{sec:introduction}

The functional form of the initial mass function (IMF) has far reaching implications in our understanding of the stellar component of galaxies, galaxy formation and evolution, and in the interpretation of observed galactic properties. It has been traditionally assumed that the IMF of extragalactic stellar populations follows the same functional form as measured in the Milky Way (MW) \citep{Salpeter,Kroupa,Chabrier03,Bastian10}. Recently, measurements of the IMF in early-type galaxies (ETGs) have called this assumption into question \citep{vDC10, Treu10, Spiniello2012, CvD12b, Capp12, LaBarbera13}. These studies have suggested that the IMF becomes increasingly `bottom-heavy,' meaning a larger fraction of low-mass ($<$1 M$_{\sun}$) stars, in ETGs with higher central velocity dispersions ($\sigma_{v}$), higher alpha element abundances ([$\alpha$/Fe]), or increasing metallicity ([Z/H]) \citep{MN15b}. This variation has also been measured in the bulges of some massive spiral galaxies \citep{Dutton2013, Parikh2018}.

There are two general methods that have been adopted to constrain the IMF of unresolved stellar populations. The first method relies on either galactic dynamics \citep[e.g.,][]{Capp12, Li17} or gravitational lensing \citep[e.g.,][]{Treu10, SLC15, Collier18} to constrain the galactic mass and therefore the stellar mass-to-light (M/L) ratio. These measured (M/L) are then compared to those derived from stellar population models to infer a mass excess that would be indicative of a variable IMF. The second method uses several surface-gravity sensitive absorption features (e.g. \nai{} 0.84 \micron, \ion{Ca}{2}{} Triplet 0.86 \micron, Wing-Ford band (FeH) 0.99 \micron) along with stellar population synthesis (SPS) to fit the IMF directly from the integrated galaxy spectrum \citep[e.g.,][]{vDC12, Spiniello2012, LaBarbera13, Kin17, Ziel16}. Both these methods have independently found evidence for IMF variations in galaxies, strengthening the overall argument for the existence of some effect. However, different studies have identified discrepancies between the IMFs measured for particular galaxies \citep{Smith14}, on the existence of IMF variation \citep{SL13, SLC15}, or the cause of the variation \citep{LaBarbera15}. Reconciling these discrepancies requires additional confirmation based on evidence from new or expanded methods.

Constraining the IMF using gravity sensitive features is enticing as it directly probes the galactic stellar populations. They help break the degeneracy arising over the relative contribution of dwarf and giant stars with similar effective temperatures to the integrated light. Absorption features, however, are highly sensitive to stellar population parameters such as age, star formation history, and elemental abundances \citep[][hereafter CvD12a]{CvD12a}. This requires either additional features, full spectral modelling, or broader spectral coverage to more reliably constrain these population parameters.

A majority of studies that have employed SPS have focused on using absorption features in the visible bands (up to 1.0 \micron). Few studies have extended further into the NIR, with notable exceptions \citep[e.g.][hereafter ASL17]{Kin17}. There are advantages to exploring the NIR, relative to the visible bands, for use in constraining the IMF, including its increased sensitivity to light from faint, low-mass dwarfs. The spectral intensity of M dwarfs, which tend to dominate the stellar mass of galaxies regardless of the shape of the IMF, peaks in the range of 0.8--1.2 \micron{} \citep{Raj13}.
In addition, a significant percentage of the light emitted by late type M dwarfs, i.e. M $\leq 0.3$ M$_{\sun}$, is emitted in the NIR.
According to the CvD12a models, there are numerous gravity-sensitive absorption features in the zJ-band (0.9 -- 1.35 \micron) that vary by 1\% or more between a Kroupa (MW-like) and bottom-heavy IMF. These features are: the Wing-Ford band (FeH) at 0.99 \micron{}, a \cai{} line at 1.13 \micron, a \nai{} line at 1.14 \micron{}, a \kai{} Doublet at 1.17 \micron{}, a \kai{} line at 1.25 \micron, and a \ali{} line at 1.31 \micron{}.

In this paper we present a pilot study on constraining the IMF of nearby ETGs using the above set of zJ-band gravity sensitive features. For this, we observed the innermost cores of two nearby ETGs. Recent spatially resolved studies of nearby ETGs have found indications that the IMF slopes in these galaxies vary radially \citep{Sarzi17, vDC16, LaB16, MN15}. These studies suggested that the bottom-heavy IMF may be localized in the cores of massive ETGs while the IMF becomes more MW-like at larger radii. Galactic cores are, therefore, ideal locations for investigating variations in the IMF. We compare our observed spectra to the latest version of the \citet{CvD12a} models \citep[][hereafter C18]{Conroy18}. Our analysis adopts two standard methods: equivalent width (EW) measurement and spectral fitting based on a Markov Chain Monte Carlo (MCMC) technique, and we conduct a detailed comparison between the results from the two methods.

This paper is organized as follows: we provide the details of our observations and data reduction procedure, the measured equivalent widths, the MCMC simulations, and discussion in \S\ref{sec:ObsData}, \S\ref{sec:EWAnalysis}, \S\ref{sec:MCMCResults}, and \S\ref{sec:discussion}, respectively, followed by a summary and our conclusions in \S\ref{sec:conclusions}.

%%%%%%%%%%%%%%%%%%%
\section{OBSERVATIONS AND DATA REDUCTION}\label{sec:ObsData}
We chose two galaxies, M85 and M87, from the \citet{vDC12} (hereafter vDC12) target sample that were shown to have highly contrasting, central IMFs. Selected parameters for the targets can be found in Table \ref{tab:properties}. \citet{CvD12b} (hereafter CvD12b) concluded that M85 has a MW-like IMF while M87 has a Salpeter type. M87 has also been the subject of IMF variation analysis using SPS in ASL17 and \citet{Sarzi17} and dynamical modelling in \citet{Oldham18}, all of which have found a Salpeter-like or steeper IMF. Furthermore, these two galaxies are on opposite ends of both the IMF--$\sigma_{v}$ and IMF--[$\alpha$/Fe] distributions in CvD12b. They are therefore suited as exemplary targets for observational analysis of stellar populations with contrasting IMFs. 

\subsection{Observations}
We carried out seeing-limited, integral field spectroscopic observations of M85 and M87 with the Near-Infrared Integral Field Spectrometer \citep[NIFS;][]{NIFS} on the 8-meter Gemini North Telescope as part of the \textit{GN--2015A--Q--47} programme. M85 was observed on May 8th, 27th, and 30th, 2015 while M87 was observed on June 2nd and 5th, 2015. The observations were made using the zJ--filter and both the z-band (0.94--1.15 \micron) and the J-band gratings (1.15--1.33 \micron), respectively. This provided an effective spectral coverage of 0.94--1.33 \micron. These bands have spectral resolving powers in the range of $\sim$5000--6000. The NIFS integral field size was $\sim$3\arcsec~$\times$~3\arcsec{} mapped by 29 image slices, and a pixel scale of 0.\arcsec103 and 0.\arcsec04 across and along the slices, respectively. 

The targets were observed with a repeated target--sky--target nodding sequence. The z- and J-bands for M85 as well as the J-band for M87 were observed for $8\times300$s exposures and $4\times300$s sky exposures. M87 was observed for $6\times300$s exposures and $3\times300$s sky exposures in the z-band. During each observing night, we observed an A0V standard star at a similar airmass as the target galaxy for use in removing telluric features from the galaxy spectrum. 

Our analysis relied on using NIFS as a spectrographic `light-bucket' for obtaining an integrated spectrum with a high signal-to-noise ratio (S/N). It is critical to have a high S/N spectrum so as to resolve the minute differences in flux (1-2\%) caused by IMF variation in each of the absorption features. 

\begin{deluxetable*}{ccccccccc}
\tablecaption{Selected galaxy parameters of the target sample.\label{tab:properties}}
\tablehead{
\colhead{Galaxy} & \colhead{RA} & \colhead{DEC} & \colhead{$T_{exp}$ (min)} & \colhead{z} & 
\colhead{$\sigma$ (\kms)} & \colhead{[Mg/Fe]} & \colhead{[Fe/H]} & \colhead{\alphak}
}
\colnumbers
\startdata
M85 (NGC~4382) & 12:25:24.1 & +18:11:29 & 40 (Z), 40 (J) & 0.0024 & 170 & 0.11 & -0.02 & 0.63 \\ 
M87 (NGC~4486) & 12:30:49.4 & +12:23:28 & 30 (Z), 40 (J) & 0.0043 & 370 & 0.33 & -0.16 & 1.90 \\
\enddata
\tablecomments{(1) Target galaxy (2) Right ascension (3) Declination (4) Total exposure time (5 min exposures) (5) Redshift from \citet{Smith2000} (6) Velocity dispersion averaged in the central 3\arcsec{} from \citet{Emsellem04} (7)--(9) From CvD12b; \alphak{} = (M/L)$_{K}$/(M/L)$_{K,MW}$ is the `IMF-mismatch' parameter (see \S\ref{sec:MLR}). This parameter measures the 'bottom-heaviness' of the IMF.}
\end{deluxetable*} 

\subsection{Data Reduction}\label{sec:reduction}
The obtained NIFS data were primarily reduced using a suite of \textsc{python}/\textsc{iraf} reduction scripts provided by the standard Gemini NIFS pipeline. The pipeline scripts perform flat fielding, sky subtraction, spatial distortion correction, wavelength calibration, and spectral cube extraction. Additional data processing such as atmospheric absorption (telluric) correction and uncertainty estimation were accomplished with custom \textsc{python} routines. 

Atmospheric absorption in the galaxy spectra was corrected for using the spectrum of a A0V telluric star observed at a similar airmass as the galaxy observations. Before being applied to the galaxy spectrum, the telluric star spectrum was first corrected for intrinsic A0V star absorption features, such as the strong Paschen series lines, following the method outlined in \cite{Vacca2003}. This method employs a template Vega spectrum that is scaled to match the observed A0V star absorption features. 

In order to remove the telluric features from the galaxy spectra, the telluric standard spectrum was aligned with the respective galaxy spectrum using 7 telluric regions that appear in both spectra. This was necessary as the wavelength solutions provided by the NIFS pipeline were not precise enough for the atmospheric features to be aligned across the observed bandpass. The telluric features were then removed by dividing the galaxy spectra by the corrected telluric standard spectrum.

The \nai{} 1.14 \micron{} feature is located in a region of strong telluric absorption at the end of the NIFS z-band. The above telluric correction method was repeated on this feature in isolation so as to further reduce the effects of atmospheric absorption on the feature profile. 

M87 has a strong, central active galactic nucleus (AGN) that likely contaminates the measured absorption features \citep{Kin17,Sarzi17}. Two strong emission features were identified in the final M87 spectrum: a [\ion{S}{3}] + P~$\epsilon$ feature at $\sim$0.953 \micron{}, and a weaker [\ion{Fe}{2}] feature at $\sim$1.26 \micron. The [\ion{Fe}{2}] feature intersected with the \kai{} 1.25 \micron{} feature in the EW index measurement bands. It was removed from the M87 spectrum with a gaussian profile fit; however, the residual effects significantly reduced our confidence in the measurement of the \kai{} 1.25 \micron{} feature. We therefore excluded this feature from our analysis of the observed M87 spectrum.

In addition to the line emission, continuum emission from the AGN also likely contaminated the observed M87 spectrum. ASL17 used Hubble observations of M87 to measure the level of AGN contamination to the spectrum continuum level. They estimated an approximately 15\% contribution to the total M87 continuum level. Since the field of NIFS is almost identical to that of the KMOS field in their study, we consider the effect a 15\% continuum correction would have on our IMF measurements in \S\ref{sec:AGNCorrection}.

\begin{deluxetable}{lccc}
    \tablecolumns{4}
	\tablecaption{Equivalent width index names and definitions.\label{tab:lines}}
\tablehead{
	\colhead{Index} & \colhead{Feature (\AA)} & \colhead{Blue} & \colhead{Red} \\ \colhead{} & \colhead{} & \colhead{Continuum (\AA)} & \colhead{Continuum (\AA)}}
\startdata
FeH 0.99    & 9905--9935   & 9855--9880   & 9940--9970   \\ 
\cai{} 1.03 & 10337--10360 & 10300--10320 & 10365--10390 \\
\nai{} 1.14 & 11372--11415 & 11340--11370 & 11417--11447 \\
\kai a 1.17 & 11680--11705 & 11667--11680 & 11710--11750 \\
\kai b 1.17 & 11710--11750 & 11793--11810 & 11765--11793 \\
\kai{} 1.25 & 12460--12495 & 12555--12590 & 12505--12545 \\
\ali{} 1.31 & 13090--13113 & 13165--13175 & 13115--13165 \\
\enddata
\end{deluxetable} 

\begin{figure*}
\epsscale{1.1}
\plotone{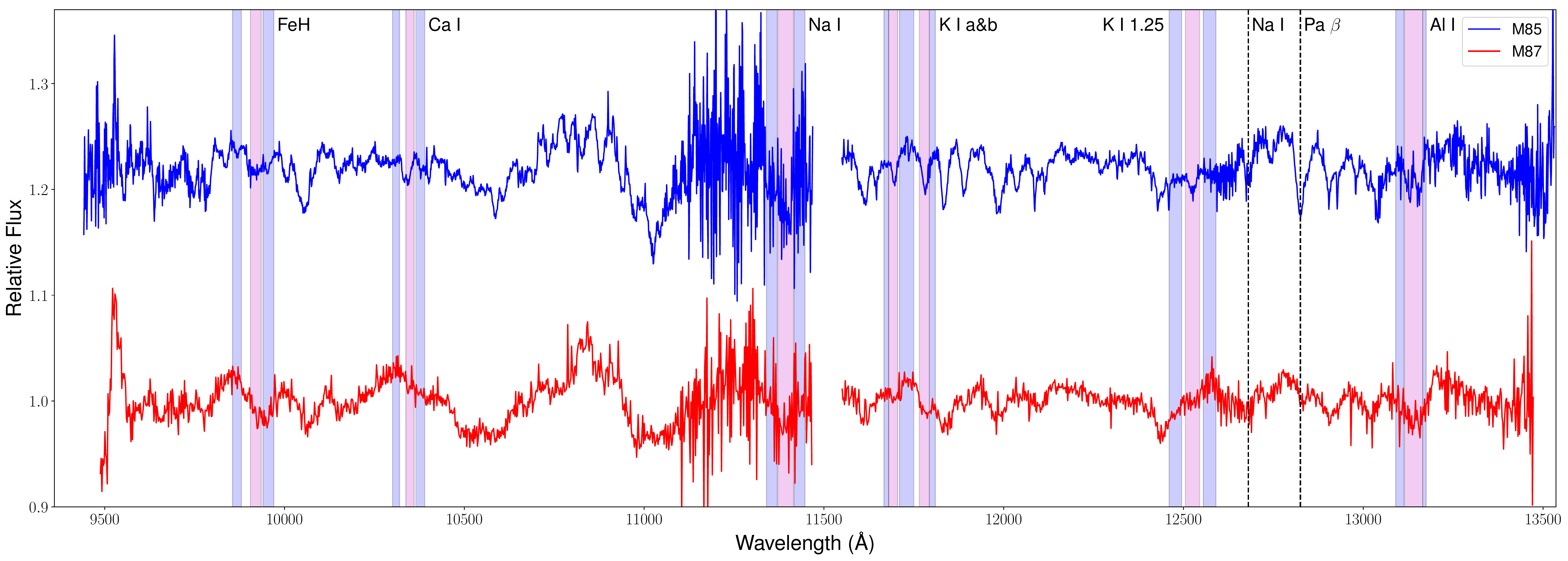}
\caption{Fully collapsed and reduced spectra for both M85 (solid blue line) and M87 (solid red line). The spectra were normalized by dividing by a high-order polynomial for clarity. The absorption features that we measure in this paper are labelled and highlighted (Table \ref{tab:lines}). The equivalent width continuum bands are shaded in blue and the measurement bands are shaded in magenta. The emission feature seen in the M87 spectrum at 9500 \AA{} is a [\ion{S}{3}] + Pa~$\epsilon$ composite feature from the central AGN. This line was not removed from the spectrum as it does not affect the measurement of any of the IMF-sensitive features.\label{fig:spectra}}
\end{figure*}

The final, telluric-corrected, integrated spectra for both galaxies are shown in Figure \ref{fig:spectra}. We removed the background continuum level by fitting the spectra with 7th order polynomials. The spectral bands that we use in the EW index analysis (see \S\ref{sec:EWAnalysis}) are shaded: the blue shaded regions are the continuum bands, and the magenta regions are the measurement bands. The spectral resolution of the M87 spectra was halved (R$\sim$3000) in order to improve the overall S/N. This was implemented by taking the mean spectral value for consecutive pairs of wavelength points. It was necessary to improve the S/N due to the lower integration time as well as the high central velocity dispersion in M87. Higher velocity dispersions reduces the strength of the spectral indices by spreading the signal over a wider bandpass, thereby requiring higher S/N measurements to accurately measure the effects of the IMF on the index strengths. The final, mean S/N per resolution element across the full zJ-band ($\sim$0.95 -- 1.35 \micron) is 71 and 61 for M85 and M87, respectively.

\section{EQUIVALENT WIDTH ANALYSIS}\label{sec:EWAnalysis}

\subsection{Equivalent Width Measurements}\label{sec:ewmeasurement}
In \S\ref{sec:introduction}, we identified seven gravity-sensitive absorption features in the zJ-band that vary by greater than 1\% between stellar populations with a bottom-heavy IMF and those with a MW-like IMF. 
The EW index definitions for these features are identical to those given in CvD12a, Kin17, and \citet{SLC12} and are listed in Table \ref{tab:lines}. 
All seven features increase in strength for stellar populations with a higher percentage of dwarf stars than giants.
Any giant star sensitive features in the zJ-band vary weakly with respect to the IMF so they were not included in the following analysis due to the high S/N required to accurately measure the effects of the IMF.

We defined fiducial models from C18 for each galaxy in order to provide a standard comparison to the measured indices.
The stellar ages and metallicities of the fiducial models closely match those measured within a $R_{e}$/8 radius for M85 (5.0 Gyr Age, [Z/H] = 0.0) and M87 (13.5 Gyr Age, [Z/H] = 0.2) in \citet{ATLAS3D30}.
The observed NIFS fields for both galaxies were fully contained within a $R_{e}/8$ radius and accounts for 18\% and 32\% of the $R_{e}/8$ radii for M85 and M87, respectively \citep{Kraj2013}.

\begin{figure*}[h] 
%\epsscale{\textwidth}
\plotone{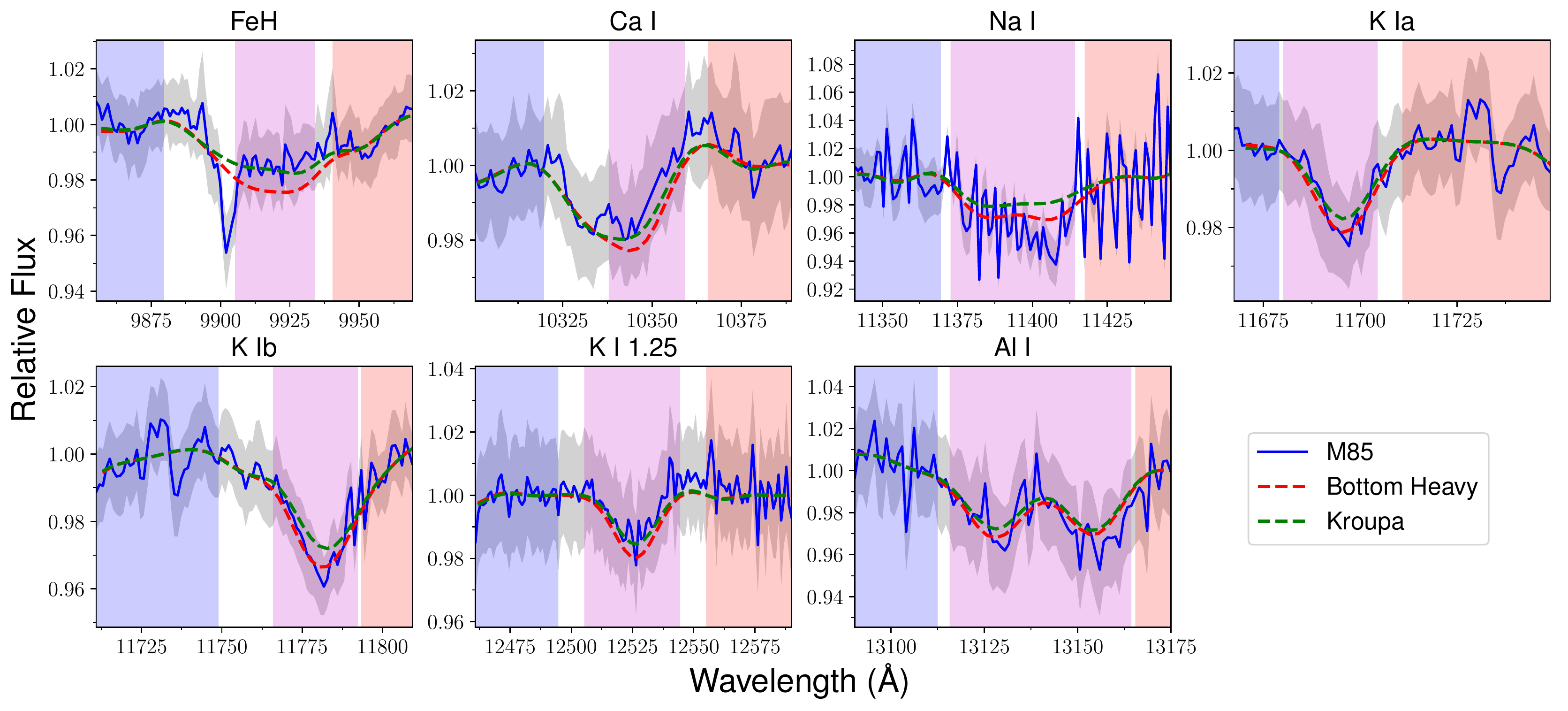}
\caption{The seven observed IMF-sensitive absorption features from the observed M85 spectrum (solid blue lines). The dashed lines are fiducial models with a stellar age of 5.0 Gyr and solar metallicity. Models with Kroupa and bottom-heavy IMFs are seen as the green and red dashed lines, respectively. The model spectra have been been broadened to match the central velocity dispersion of M85. The blue and red shaded areas correspond to the EW bandpass regions and the magenta shaded areas corresponds to the feature index measurement region (see Table \ref{tab:lines}). The grey shaded regions correspond to a 68.3\% confidence level in the spectral intensity.\label{fig:M85Lines}}
\end{figure*}

\begin{figure*}[h] 
%\epsscale{\textwidth}
\plotone{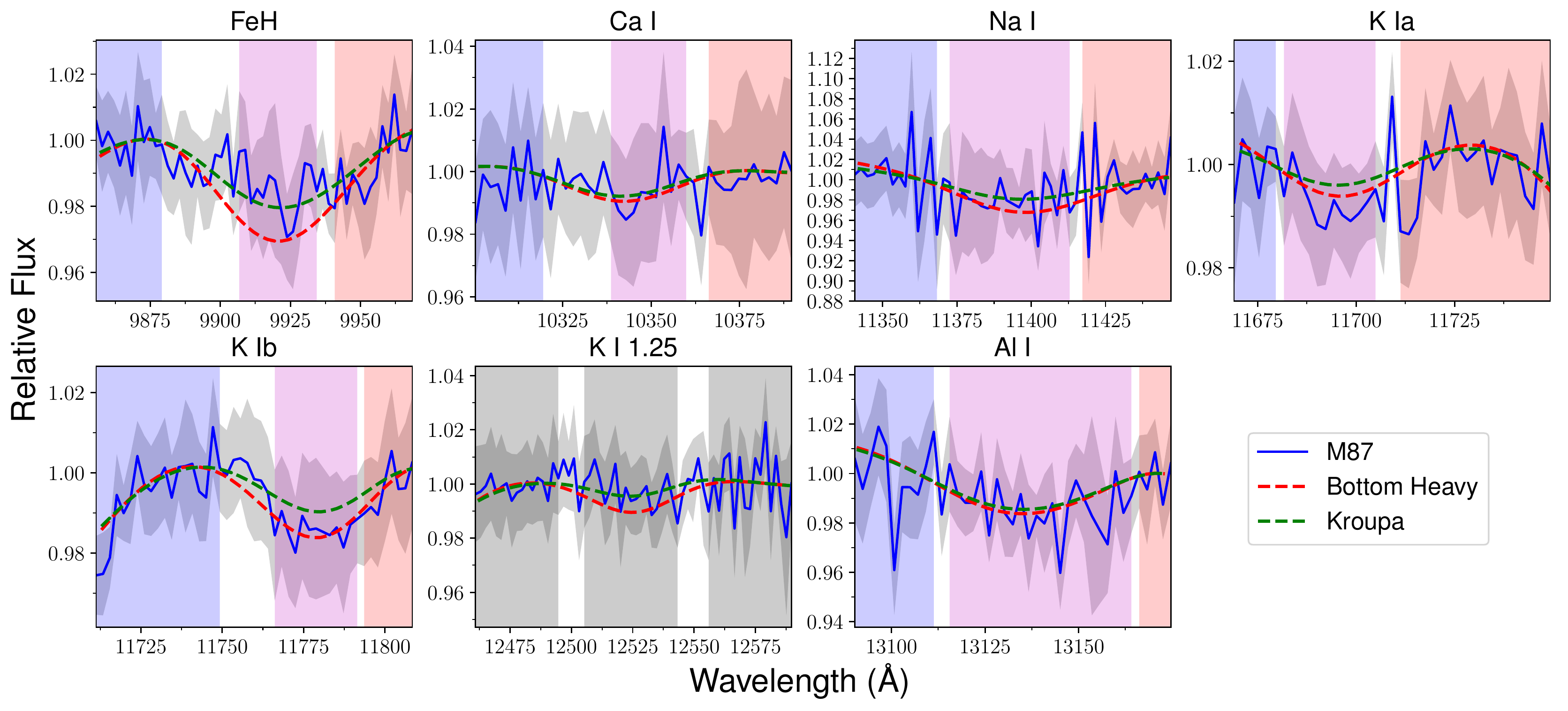}
\caption{Same as Figure \ref{fig:M85Lines}. The fiducial models for M87 have a stellar population age of 13.5 Gyr and [Z/H] = 0.2. The EW bandpass regions for the \kai{} 1.25 feature are grey as this feature was excluded from our analysis of the M87 spectrum.\label{fig:M87Lines}}
\end{figure*}

We show a close--up view of the seven observed absorption features in Figures \ref{fig:M85Lines} and \ref{fig:M87Lines}.
The observed spectra are represented by the solid blue lines while fiducial models with a Kroupa and bottom-heavy IMF are shown in green and red dashed lines, respectively.
The model spectra were broadened to match the central velocity dispersions of each galaxy (see Table \ref{tab:properties}).
The blue, red, and magenta shaded regions in both figures define the EW continuum and line bandpasses, respectively. 

We measured the EW indices of the IMF-sensitive absorption features following the method of \citet{Trager98}.
Table \ref{tab:MeasureTable} lists the measured index strengths with their estimated uncertainties and the mean S/N across the feature band for each index. 
The uncertainties were calculated based on the variance in the intensity at a given wavelength in the set of 6-8 target exposures.
The S/N for the z-band EW in M87 are lower due to the shorter total integration time in that filter relative to the surface brightness of the galaxy.

\subsection{Adopted Stellar Population Models}
In the following analysis, we make use of two groups of synthetic simple stellar population models from C18: one group that varies the IMF while holding other stellar population parameters (e.g. age, metallicity) constant, and another group of `response functions' for individual elemental abundances (e.g. [Na/H], [Fe/H], [Ca/H], [K/H], etc).
These models are computed by combining observed stellar spectra from the MIST \citep{Choi2016} and Extended IRTF \citep{Villaume2017} spectral libraries, with the theoretical response functions.
Both groups of models are subdivided into sets with fixed metallicities ranging from [Z/H] = --1.5 to +0.2 and stellar ages from 1.0 to 13.5 Gyr. 
The former group of models are used to characterize the IMFs of M85 and M87 and the latter to infer the effects of possible abundance variations. 

The group of models with variable IMF slopes follows a broken power law for the IMF functional form. The power law is split into three mass intervals with exponents ($x_{i}$):
\begin{eqnarray}
    x_{1}:& &0.08 \leq M/M_{\sun} < 0.5 \nonumber\\
    x_{2}:& &0.5 \leq M/M_{\sun} < 1.0\\
    x_{3}:& &1.0 \leq M/M_{\sun} \nonumber
\end{eqnarray}
The \imfi{} and \imfii{} exponent values range from 0.5 to 3.5 in steps of 0.2. The IMF slope for stellar masses above 1.0 $M/M_{\sun}$, $x_{3}$, is static and identical to the Kroupa IMF value of 2.3. We refer to `X' as a single slope IMF exponent, equivalent to an IMF where \imfi{} and \imfii{} are equal, i.e. a Salpeter IMF is defined as X = 2.3, and a bottom-heavy IMF as X $\geq$ 3.0. 

\subsection{Constraints on the IMF from Individual Line Indices}\label{sec:lineanalysis}
\begin{figure*}
\epsscale{1.0}
\plotone{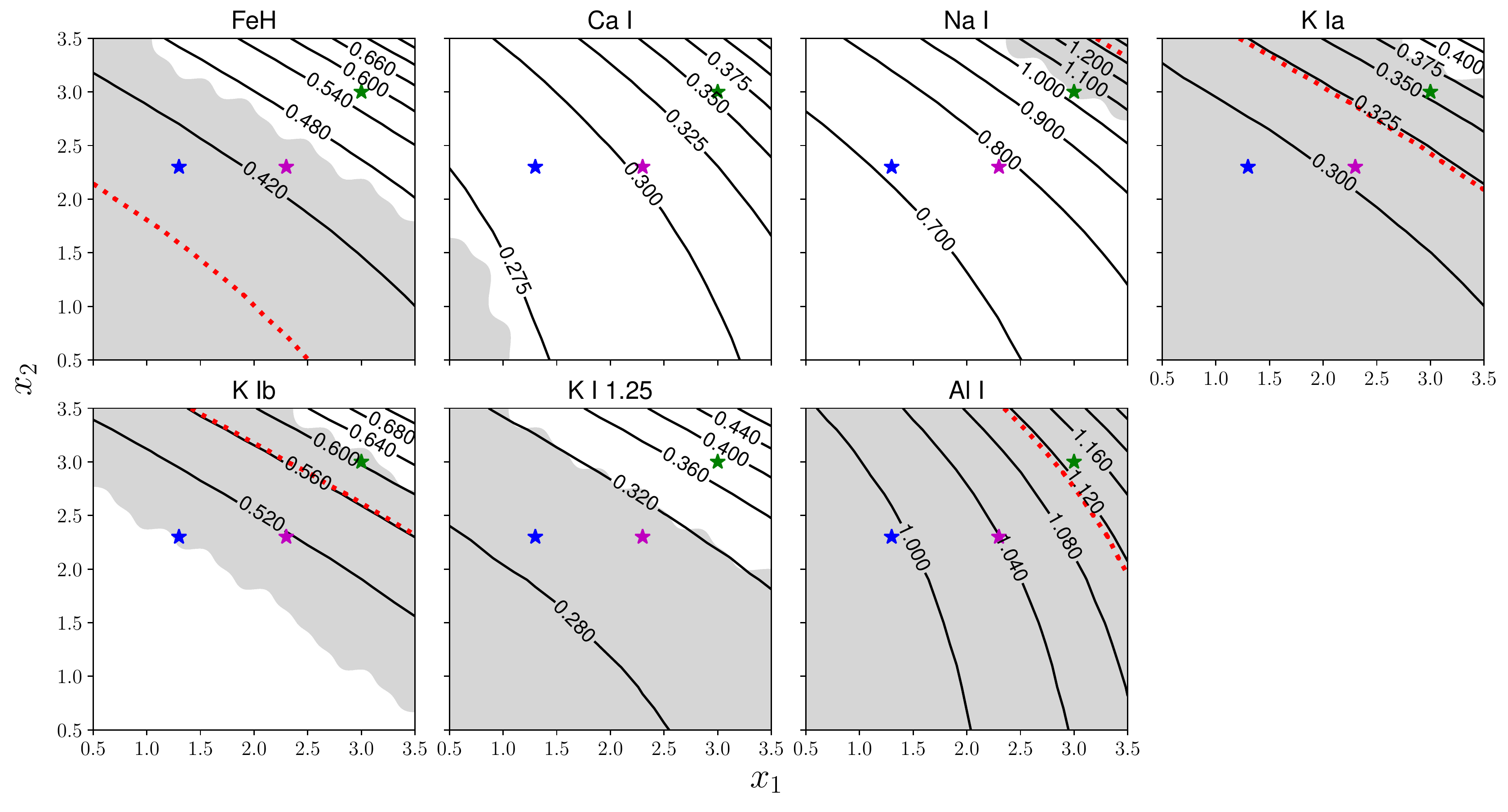}
\caption{Model EW index strengths for each of the seven absorption features measured in the M85 spectrum, relative to the IMF power-law exponents, \imfi{} and \imfii, measured from the fiducial models. The black and red contours represent lines of constant index strength and the observed index strength, respectively. Kroupa, Salpeter, and bottom-heavy IMFs are marked as blue, magenta, and green stars, respectively. The grey regions represent the the index--space covered by the measured index strength within a 68.3\% confidence level. The strengths of individual indices are inconclusive regarding the best fit IMF slopes.\label{M85Model}}
\end{figure*}

\begin{figure*}
\epsscale{1.0}
\plotone{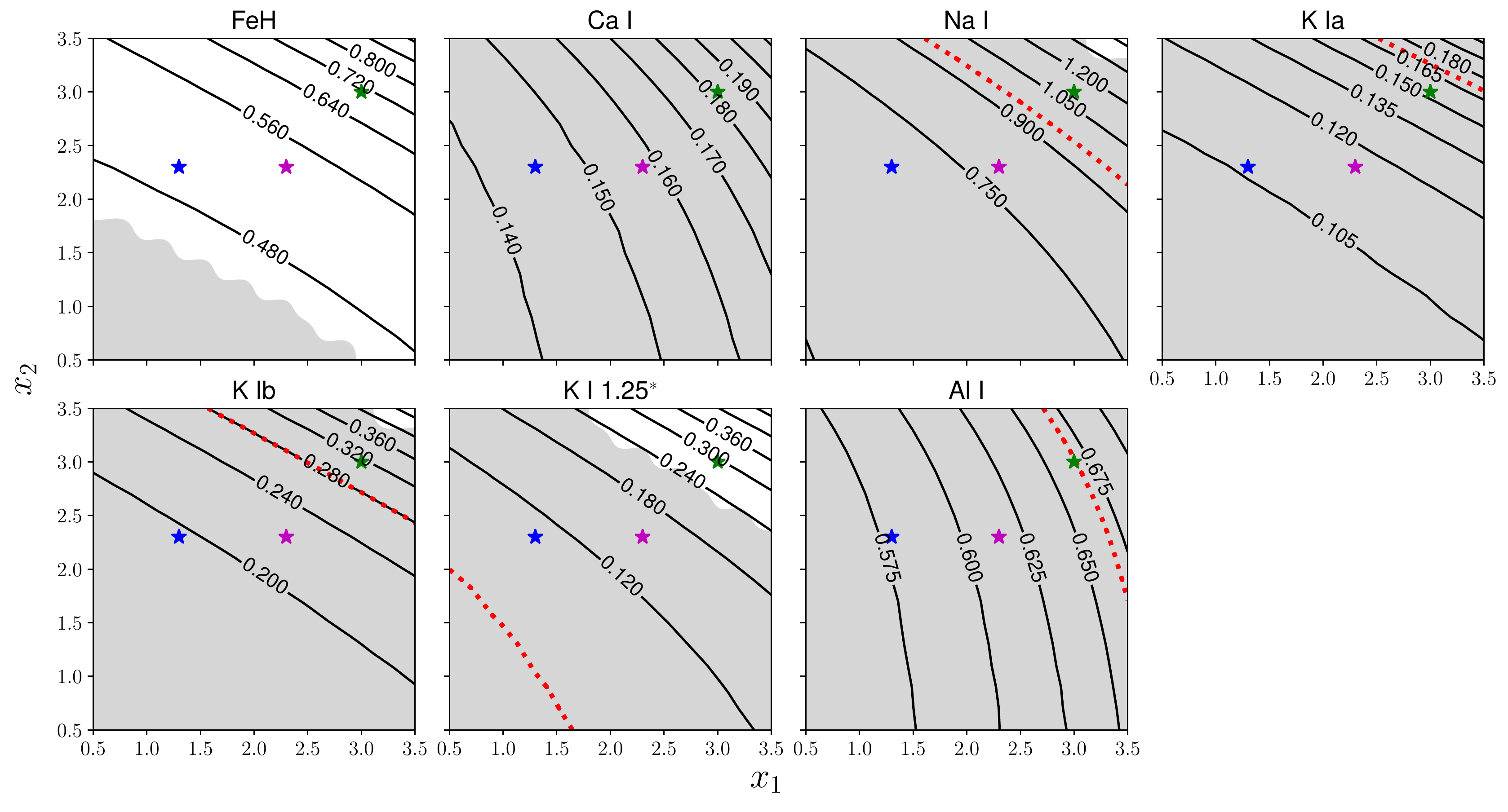}
\caption{Same as Figure \ref{M85Model} for the indices measured from the M87 spectra. Note that the uncertainties are large enough to cover the majority of the index--space for many of the M87 absorption features. This makes drawing a conclusion regarding the best-fit IMF slopes difficult with the index strengths. (*The KI 1.25 feature is presented for comparison purposes but is not included in the analysis below.)\label{M87Model}}
\end{figure*}

We compare the observed EW indices for both M85 and M87 to those measured from the broadened fiducial models in Figures \ref{M85Model} and \ref{M87Model}.
The indices are plotted as a function of the two low-mass IMF exponents, \imfi{} and \imfii, with the black and red contours marking lines of constant model index and the observed index, respectively.
Also included for reference are Kroupa, Salpeter, and bottom-heavy (X = 3.0) IMFs shown as the blue, magenta, and green star symbols, respectively. 

In Figures \ref{M85Model} and \ref{M87Model}, the observed index strength of some of the features, e.g., FeH and \nai{} 1.14 \micron, show discrepancies in their favored IMF slopes, making it difficult draw consistent conclusions regarding the IMF slopes in either galaxy.  
A likely explanation for these discrepancies is the strong degeneracy between the effects of the IMF and other stellar population parameters, such as elemental abundances and stellar ages, on the measured index strengths \citep{CvD12a}.
This indicates that a simple analysis based on general stellar population parameters available in the literature is inadequate for constraining the IMF; consequently, we need a more rigorous analysis that takes the effect of the varying abundance ratios into account to more effectively constrain the IMF.
Below, we conduct an investigation into the effects of varying elemental abundances on the measured index strengths for each observed feature.
This investigation is used to interpret the results of a more thorough study of the IMFs of the observed galaxies in \S\ref{sec:MCMCResults}.

\begin{deluxetable}{lCcCc}
\tablecaption{Measured IMF-sensitive index strengths.\label{tab:MeasureTable}}
\tablehead{
	\colhead{Line} & \colhead{EW} & \colhead{S/N\tablenotemark{a}} & \colhead{EW} & \colhead{S/N\tablenotemark{a}} \\
	\colhead{} & \colhead{M85} & \colhead{M85} & \colhead{M87} & \colhead{M87}
}
\startdata
FeH 0.99 & 0.39\pm0.08 & 81  & 0.31\pm0.11 & 67 \\ 
\cai{} 1.03 & 0.20\pm0.07 & 73  & 0.06\pm0.15 & 50 \\
\nai{} 1.14 & 1.33\pm0.24 & 53  & 0.95\pm0.50 & 28 \\
\kai a 1.17 & 0.32\pm0.07 & 84  & 0.17\pm0.09 & 95 \\
\kai b 1.17 & 0.56\pm0.06 & 100 & 0.28\pm0.14 & 64 \\
\kai{} 1.25 & 0.21\pm0.11 & 67  & \multicolumn{2}{c}{--} \\
\ali{} 1.31 & 1.11\pm0.18 & 45  & 0.67\pm0.25 & 59 \\
\enddata
\tablenotetext{a}{Mean S/N within the index feature band defined in Table \ref{tab:lines}.}
\end{deluxetable}

\subsubsection{Wing-Ford Band (FeH 0.99 \micron)}\label{sec:FeH}
The FeH 0.99 \micron{} feature is widely included in IMF variation studies \citep[e.g.][]{vDC12, Kin17, Vaughan2018} due to its low-sensitivities to a stellar population's age and $\alpha$--element abundance \citep{CvD12a}. 
It is the only molecular feature in the measured set of seven features. 
According to the C18 models, the strength of this index is primarily sensitive to [Fe/H] and has a negative relationship with [Na/H]. 
The latter relationship originates from the role of Na as a major electron donor in cool stars.
A high Na abundance encourages the dissociation of the FeH molecule in stellar atmospheres \citep[CvD12a; ASL17;][]{SLC12, Vaughan2018}.

The observed index strength of the FeH 0.99 \micron{} feature is 0.39~$\pm$~0.08 \AA{} in M85 and 0.31~$\pm$~0.11 \AA{} in M87.
For M85, this index strength is consistent (within a 68.3\% confidence level) with IMF slopes similar to a Kroupa-like IMF as predicted in CvD12b.
In M87, the FeH index is very weak relative to what is expected for a stellar population with a bottom-heavy IMF as suggested by CvD12b and others.

One possible cause for the low FeH index strength in M87 is an underlying low [Fe/H] and/or a high [Na/H]. 
CvD12b measured a [Fe/H] = --0.16 dex in the core of M87 and \citet{Sarzi17} measured a high central [Na/H] = +0.7 dex . 
We find that the C18 models predict that the FeH index is reduced by approximately 13\% at [Fe/H] = --0.16 dex and by 25\% at [Na/H] = +0.7 dex. 
Correcting the index strength for these abundance effects increases the measured index to 0.43~$\pm$~0.15 \AA{} which would instead be marginally consistent with a Salpeter-like IMF.
Other possible factors influencing the strength of the FeH index include increased uncertainties arising from measuring this index for a galaxy with high $\sigma_{v}$ \citep{McConnell2016}, or spectral contamination from the central AGN.

\subsubsection{Calcium Feature (\cai{} 1.03 \micron)}
The \cai{} 1.03 \micron{} feature is the weakest IMF-sensitive feature in our set. 
The strength of the \cai{} 1.03 \micron{} index is strongly sensitive to [Ca/H] but it is not affected by variations in [Na/H], unlike other IMF indicators in both the visible and NIR bands \citep{SLC12}.
We measured a \cai{} index strength of 0.20~$\pm$~0.07 \AA{} in M85 and 0.06~$\pm$~0.15 \AA{} in M87.
These indices are very weak in both galaxies, particularly in M87, relative to the expected index strengths in the fiducial models (Figures \ref{M85Model} and \ref{M87Model}).
The low \cai{} index strength in M87 may be the result of a low Ca abundance, as found in ASL17 for a set of similarly massive ETGs.

\subsubsection{Sodium Feature (\nai{} 1.14 \micron)} \label{sec:sodium}
The \nai{} 1.14 \micron{} feature is the most dwarf star sensitive feature in the zJ-band according to the C18 models.
In addition, the models indicate that the index strength has a strong, negative relationship with [Fe/H], similar to the relationship between FeH and [Na/H].

The measurement of this index is complicated by its location within a densely populated band of telluric water absorption lines for low-redshift galaxies.

We measured a very strong \nai{} 1.14 \micron{} index in both observed galaxies, 1.33~$\pm$~0.24 \AA{} in M85 and 0.95~$\pm$~0.50 \AA{} in M87, relative to the expected index strengths from the fiducial models. 
In M85 and M87, the index strength is consistent with a `bottom heavy' IMF and/or a stellar population with a significantly enhanced Na abundance. 

In M85, the measured \nai{} 1.14 \micron{} index strength is greater than expected for a MW-like IMF as measured in CvD12b.
Reconciling this index strength with the fiducial model strength for a Kroupa IMF (0.72 \AA) would require either a significantly enhanced [Na/H], a higher [Z/H], a more bottom-heavy IMF, or a combination of these effects. 
If we compare the measured index strength with a stellar population model with [Z/H] = +0.2 dex instead of solar metallicity, the index strength is instead marginally consistent with a Kroupa IMF.

Our measured \nai{} 1.14 \micron{} index in M87 is consistent with a Salpeter-like IMF, albeit with very high uncertainties (Figure \ref{M87Model}).
As previously mentioned, the central [Na/H] in M87 has been measured to be significantly greater ([Na/H] $>$ 0.7 dex) than the abundance assumed in our fiducial model ([Z/H] = 0.2 dex).
Our measurements are similar to the results of \citet{Smith2015} who found this index to be extremely strong in M87 relative to stellar population models in massive ETGs. 
\citet{Smith2015} also found that the CvD12a models still under-predict the \nai{} 1.14 \micron{} index despite a close fit to the overall spectrum and a predicted Salpeter-like IMF. 

\subsubsection{Potassium \kai{} Features}
We use two \kai{} features in our NIFS data for constraining the IMF of M85:
a doublet feature (\kai a and \kai b) at 1.17 \micron{} and a feature at 1.25 \micron.
The index strengths of the doublet feature are strong in M85, suggesting the IMF slopes of M85 are similar to a Salpeter IMF. The strength of the 1.25 \micron{} index, on the other hand, favors a more MW-like IMF than Salpeter. Note that in the doublet, the \kai b index provides tighter constraints on the IMF than the \kai a index as the former is more sensitive the changes in the IMF slope than the latter.

This discrepancy in the favored IMF slopes of the two \kai{} features is seen to a lesser degree in ASL17 where the measured central index strengths of the doublet are consistent with steeper IMFs than the 1.25 \micron{} index. 
This effect may be related to the high sensitivity of the 1.25 \micron{} feature to variations in [$\alpha$/H] as the strength of this feature can be dramatically weakened as [$\alpha$/H] increases due to changes in the local continuum behaviour (ASL17).
We find that the discrepancy in the favored IMF slopes from the two \kai{} features is reduced if we instead adopt a [Z/H] = +0.2 dex model. 
This is suggestive that an increased metallicity, or an increased K abundance, may also be responsible for the discrepancies (see ASL17).

For M87, the strengths of the \kai{} doublet indices both favour Salpeter-like IMFs which is consistent with the IMFs observed in the center of the galaxy by CvD12b.
As mentioned in \S\ref{sec:reduction}, we excluded the \kai{} 1.25 \micron{} line in our analysis of M87 due to AGN effects. 

\subsubsection{Aluminum Feature (1.31 \micron{})}
We measure an \ali{} feature at 1.31 \micron{} that has not been well explored for use in constraining the IMF slopes of integrated stellar populations.
This feature was first identified as IMF-sensitive in CvD12a, and ASL17 included measurements of the index strength in their study of the IMF of massive ETGs. 
We measure an \ali{} index strength of 1.11~\plmn~0.18 \AA{} and 0.67~\plmn~0.25 \AA{} in M85 and M87, respectively. 
Neither of these index strengths favour a particular IMF due to the large uncertainties which cover the entire \imfi--\imfii{} index space for this feature.

\subsection{Comparison to Reported Index Strengths}
ASL17 measured the index strengths of all seven features listed in Table \ref{tab:lines} for M87.
In order to compare our index strengths to those measured in ASL17 we scale our measurements to a common $\sigma_{v}$ = 230 \kms.
ASL17 did not report central index strengths for FeH and \cai{} 1.03 \micron{} so we instead compare to their measurements at $R_{e}$/3.
We find that our index strengths are generally consistent with those in ASL17 with the exception of the \nai{} 1.14 \micron{} index.
We measure a scaled index of $1.24\pm0.65$ \AA{} compared to their index strength of $2.65\pm0.26$ \AA.
This discrepancy is likely a result of the lower S/N of our z-band spectrum for M87 as well as complications with the index measurement due to the dense telluric absorption around this feature. 

\citet{SLC12} measured the \cai{} 1.03 \micron{} index strength in a sample of Coma cluster ETGs with similar $\sigma_{v}$ to M85 ($>$100 \kms).
Our measured strength in M85 is consistent with their mean measurement of 0.279~$\pm$~0.024 \AA. 
\citet{Baldwin2018} measured a \nai{} 1.14 \micron{} index strength of approximately 1.0 in low $\sigma_{v}$ ETGs, which is also consistent with the strength of our measured index in M85. 

\section{SPECTRAL FITTING ANALYSIS}\label{sec:MCMCResults}
We now describe our method for fitting the observed spectra of M85 and M87 to the C18 models.
We construct a set of adjusted models to account for abundance ratios variations in Na, Fe, Ca, and K by scaling the C18 models with fixed ages and [Z/H] by the theoretical elemental response functions.
Specifically, we fit these adjusted C18 models to the observed feature spectral bands defined in Table \ref{tab:lines} while allowing for variable stellar population parameters: the age, [Z/H], IMF slopes, and the aforementioned elemental abundances.
We fit the feature spectral bands to the models, in contrast to fitting broader spectral bandpasses as in CvD12b, in order to characterize the effectiveness of this particular set of features for constraining the IMF. In addition, fitting the feature bands allows for a comparison with the similar extragalactic IMF analysis in the NIR presented in ASL17.

\subsection{Additional Features}\label{sec:additionalfeatures}

\begin{deluxetable}{lccc}
    \tablecolumns{4}
	\tablecaption{Additional line index bandpass definitions.\label{tab:newlines}}
\tablehead{
	\colhead{Index} & \colhead{Feature (\AA)} & \colhead{Blue} & \colhead{Red} \\ \colhead{} & \colhead{} & \colhead{Continuum (\AA)} & \colhead{Continuum (\AA)}}
\startdata
\ion{Na}{1} 1.27 & 12670--12690 & 12648--12660 & 12700--12720 \\ 
\pabeta & 12810--12840 & 12780--12800  & 12855-12880  \\
\enddata
\end{deluxetable}

The seven features discussed in this paper are weakly sensitive to stellar population ages in the range of 7--13.5 Gyr (Kin17).
The index strengths vary by a much greater amount ($\leq$20\%) for younger stellar populations in the range of 3--5 Gyr.
As a consequence of this stronger age--index strength sensitivity for younger stellar ages, in addition to previous measurements of a young $\sim$4 Gyr stellar population in the core of M85 \citep{ATLAS3D30, Ko2018}, we find it necessary to include the stellar population age in our model parameters.
In order to provide better constraints on the stellar age, we define a new EW index for the Paschen~$\beta$ (\pabeta) feature at 1.28 \micron{}.
The \pabeta{} feature is highly sensitive to the stellar population age, similar to the classical Lick H~$\beta{}$ feature employed to constrain stellar ages in stellar population studies in the visible bands \citep[e.g.][]{ATLAS3D30}.

Furthermore, many of the features in Table \ref{tab:lines} are sensitive to variations in the Na abundance (see \S\ref{sec:lineanalysis}).
The Na abundance is primarily constrained by the \naline{} feature, which is the most IMF-sensitive of the seven features.
Due to the importance of [Na/H] in constraining the IMF slopes in this analysis, we define a new EW index for the \nai{} feature at 1.27 \micron.
The \newnaline{} feature provides an independent constraint on the Na abundance that is not sensitive to variations in the IMF slopes \citep{Smith2015}.

\begin{figure}
\epsscale{1.1}
\plotone{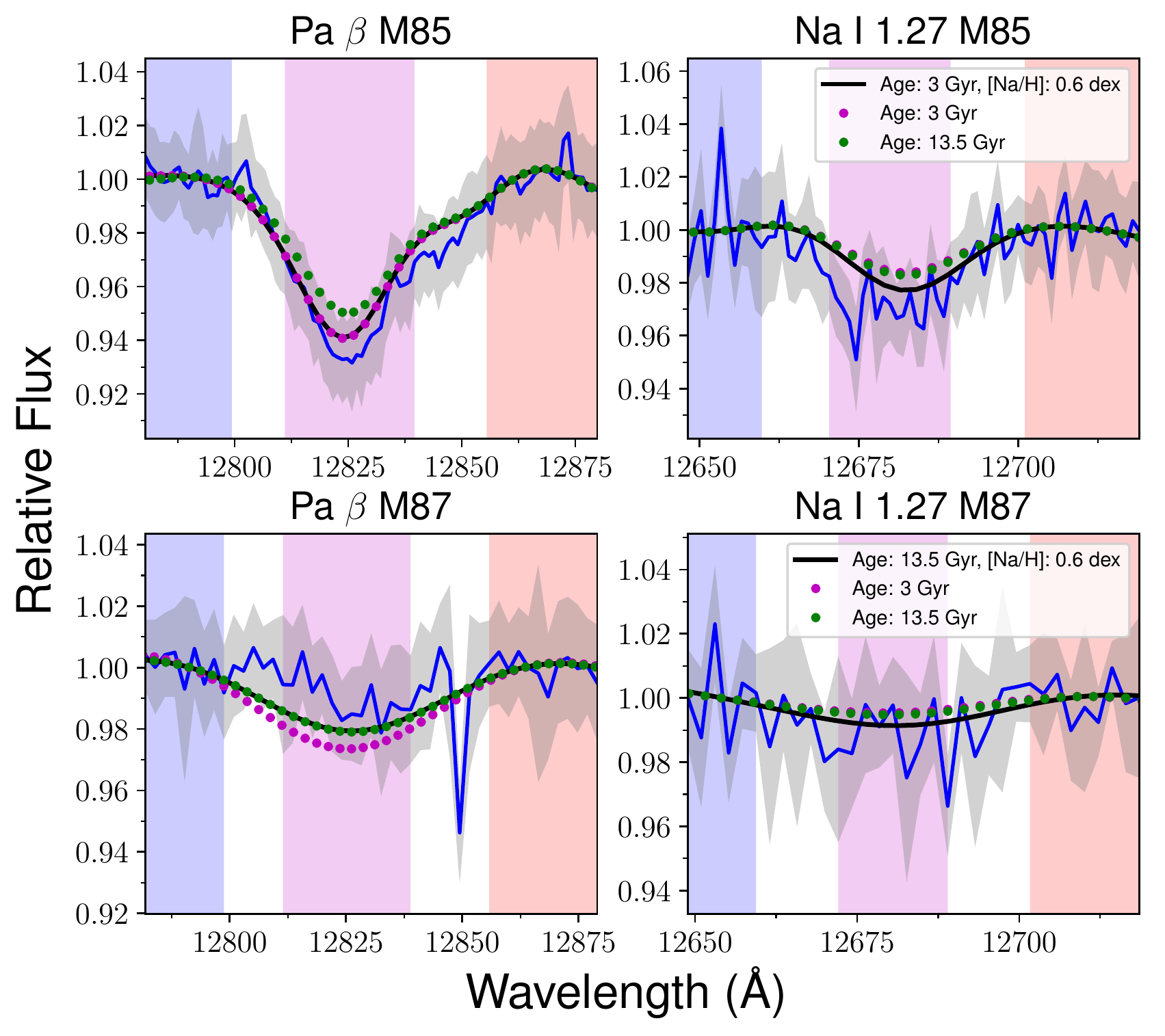}
\caption{The \pabeta{} and \newnaline{} features from the observed M85 (top row) and M87 (bottom row) spectra. The observed spectrum (solid blue line) is compared to model spectra for a stellar population with 3 and 13.5 Gyr ages (dotted lines) and one with an enhanced [Na/H] = 0.6 dex (black line). The [Na/H] enhanced model has a stellar age of 3 Gyr for M85 and 13.5 Gyr for M87. The model M85 spectra assume a Kroupa IMF while the model M87 spectra assume a bottom-heavy IMF. The model spectra have been broadened to match the velocity dispersion of each galaxy. The grey shaded regions correspond to a 68.3\% confidence level in the spectral intensity. The vertical shaded regions correspond to the EW index measurement bands as in Figure \ref{fig:M85Lines}.\label{fig:NewLines}}
\end{figure}

The EW index measurement bands for both features are defined in Table \ref{tab:newlines}.
The spectral bands are designed to have similar band widths (20--30 \AA) and total widths ($\sim$100 \AA) as the seven features in Table \ref{tab:lines}.
Profiles of the observed \pabeta{} and \newnaline{} features for both galaxies can be seen in Figure \ref{fig:NewLines} alongside stellar population models with ages of 3 and 13.5 Gyr and [Na/H] = 0.6 for illustration purposes.
The models in the figure assume either Kroupa or bottom-heavy (X = 3.0) IMFs for the M85 and M87 spectra, respectively, although neither feature is significantly sensitive to the IMF slopes.
The \pabeta{} feature, in the left column, is clearly sensitive to the stellar population age with the observed M85 and M87 spectra being better described by models with younger and older stellar populations, respectively. 
In the right column, the \newnaline{} feature is highly sensitive to the Na abundance with the observed feature profiles for both galaxies more closely aligned with the Na enhanced models.
Furthermore, \pabeta{} has a minor sensitivity to [Ca/H] while both features have minor sensitivities to [Fe/H].

\subsection{MCMC Model Overview}\label{sec:MCMCModelOverview}
The model parameters are estimated with the publicly available \textit{emcee} routine \citep{emcee}, a Markov-Chain Monte-Carlo ensemble sampler that characterizes the posterior probability distributions of a particular model provided observed data and uncertainties.
Our implementation of this routine maximizes a log-likelihood function,

\begin{equation}\label{eq:likelihood}
	ln\ p(D|\theta,\sigma) = -\frac{1}{2} \sum_{i}^{}\Big[\frac{D_{i} - M_{i}(\theta)}{\sigma_{i}}\Big]^{2}
\end{equation}

\noindent where $D_{i}$ is the observed spectrum at the $i_{th}$ wavelength element, $\theta$ are the input model parameter values listed in Table \ref{tab:priors}, $M_{i}(\theta)$ is the adjusted C18 model, and $\sigma_{i}$ is the measured uncertainty in the spectral intensity.

\begin{deluxetable}{ll}
\tablewidth{1.0\textwidth}
\tablecaption{Model Parameter priors for the MCMC ensemble sampler.\label{tab:priors}}
\tablehead{
\colhead{Parameter} & \colhead{Prior Limits}}
\startdata
~Age    & { }1.0--13.5 \\{}
[Z/H]   & --0.25--0.2 \\{}
 \imfi  & { }0.5--3.5\tablenotemark{a} \\{}
 \imfii & { }0.5--3.5\tablenotemark{a} \\{}
[Na/H]  & --0.5--0.9 \\{}
[K/H]   & --0.5--0.5 \\{}
[Ca/H]  & --0.5--0.5 \\{}
[Fe/H]  & --0.5--0.5 \\
\enddata
\tablenotetext{a}{The IMF slope exponents are limited to increments of 0.2 as in the Conory18 models.}
\end{deluxetable}

Each MCMC simulation consists of 512 `walkers' each of which explore the posterior distributions, p($\theta|$D,$\sigma$), of the model input parameters over 4000 steps or a total of $2.048\times10^{6}$ samples.
The final 1000 samples of each walker are kept for the following analysis while the preceding 3000 samples are discarded as a standard parameter `burn-in' phase.

The MCMC model input parameters and their priors are listed in Table \ref{tab:priors}.
We impose uniform, `top--hat', distributions for the priors of the input parameters in order to avoid any potential bias to the ensemble sampler.
The initial model parameters for each walker are chosen at random from within the these distributions.
We split the input parameters into two groups: a base set and an extended set.
The base set of parameters consists of the stellar population age, metallicity ([Z/H]), and the two IMF slopes \imfi{} and \imfii. 
In contrast, the extended set adds four elemental abundance ratios to the base set: [Na/H], [Fe/H], [Ca/H], and [K/H].
We fit the observed feature bands (Tables \ref{tab:lines} and \ref{tab:newlines}) of M85 and M87 to both sets of model parameters in order to assess the effects of fitting abundance variations on the IMF slope constraints.

\begin{deluxetable*}{lccccccccc}[ht]
\tablewidth{1.0\textwidth}
\tablecaption{Best-fit stellar population parameters for M85 and M87 including the \pabeta{} and \nai{} 1.27 \micron{} features.\label{tab:results}}
\tablehead{
\colhead{Galaxy} & \colhead{Age (Gyr)} & \colhead{[Z/H]} & \colhead{\imfi} & \colhead{\imfii} & \colhead{[Na/H]} & \colhead{[Fe/H]} & \colhead{[Ca/H]} & \colhead{[K/H]} & \colhead{$\alpha_{K}$\tablenotemark{a}}}
\startdata
M85 (Extended) & 3.65$_{-0.72}^{+1.25}$ & 0.17$_{-0.04}^{+0.02}$ & 1.98$_{-0.97}^{+0.92}$ & 1.47$_{-0.69}^{+0.98}$ & 0.67$_{-0.13}^{+0.13}$ & --0.03$_{-0.15}^{+0.10}$ & --0.13$_{-0.19}^{+0.19}$ & 0.09$_{-0.22}^{+0.21}$ & 1.26$_{-0.46}^{+0.81}$\\
M85 (Base) & 3.25$_{-0.51}^{+0.94}$ & 0.16$_{-0.05}^{+0.03}$ & 2.72$_{-0.67}^{+0.51}$ & 2.93$_{-0.69}^{+0.42}$ & \multicolumn{4}{c}{Fixed to [Z/H]} & 3.13$_{-1.20}^{+1.49}$\\
M87 (Extended) & 11.59$_{-2.39}^{+1.40}$ & 0.02$_{-0.16}^{+0.12}$ & 2.74$_{-1.04}^{+0.55}$ & 2.70$_{-1.01}^{+0.59}$ & 0.34$_{-0.50}^{+0.37}$ & --0.19$_{-0.22}^{+0.31}$ & --0.24$_{-0.19}^{+0.34}$ & 0.05$_{-0.35}^{+0.31}$ & 2.77$_{-1.49}^{+2.09}$\\
M87 (Base) & 11.60$_{-2.39}^{+1.41}$ & --0.03$_{-0.14}^{+0.14}$ & 2.83$_{-1.02}^{+0.49}$ & 2.75$_{-1.01}^{+0.55}$ & \multicolumn{4}{c}{Fixed to [Z/H]} & 3.02$_{-1.61}^{+2.05}$\\
\enddata
\tablenotetext{a}{Best-fit `IMF-mismatch parameter, see \S\ref{sec:MLR}}
\end{deluxetable*}

In order to calculate the log-likelihood (Equation \ref{eq:likelihood}) we first create `adjusted models' by scaling and broadening the models from C18 as a function of the input abundance ratios and the known velocity dispersion of M85 and M87 (Table \ref{tab:properties}), respectively.
The log-likelihood is then calculated by comparing the feature bands between the observed spectra and the `adjusted model.'
The feature bands are normalized with the same method detailed in \S\ref{sec:ewmeasurement} for calculating the EW index strengths.

We account for variable abundance ratios in our adjusted models by linearly interpolating the elemental response functions along the spectral and abundance ratio ([X/H]) axes. This method is a standard assumption when modelling near--solar abundance variations \citep{CvD12b, Kin17}.
C18 provides spectral response functions at \plmn0.3 dex for [Fe/H], [Ca/H], and [K/H], while they do so for [Na/H] in the range of --0.3 -- +0.9 dex with an increment of 0.3 dex.
The extended abundance ratio coverage for Na is due to its critical role in determining the depth of many of the IMF-sensitive absorption features (see \S\ref{sec:sodium}).
We increase the MCMC prior limits for the abundance ratios from $\pm$0.3 dex to $\pm$0.5 dex to conduct more thorough investigation into the effect of the abundance ratios on the IMF slopes.

The stellar population age and metallicity parameters are accounted for in a similar method as the abundance ratios by linearly interpolating between models with fixed ages and metallicities.

\begin{figure*}
\epsscale{1.1}
\plotone{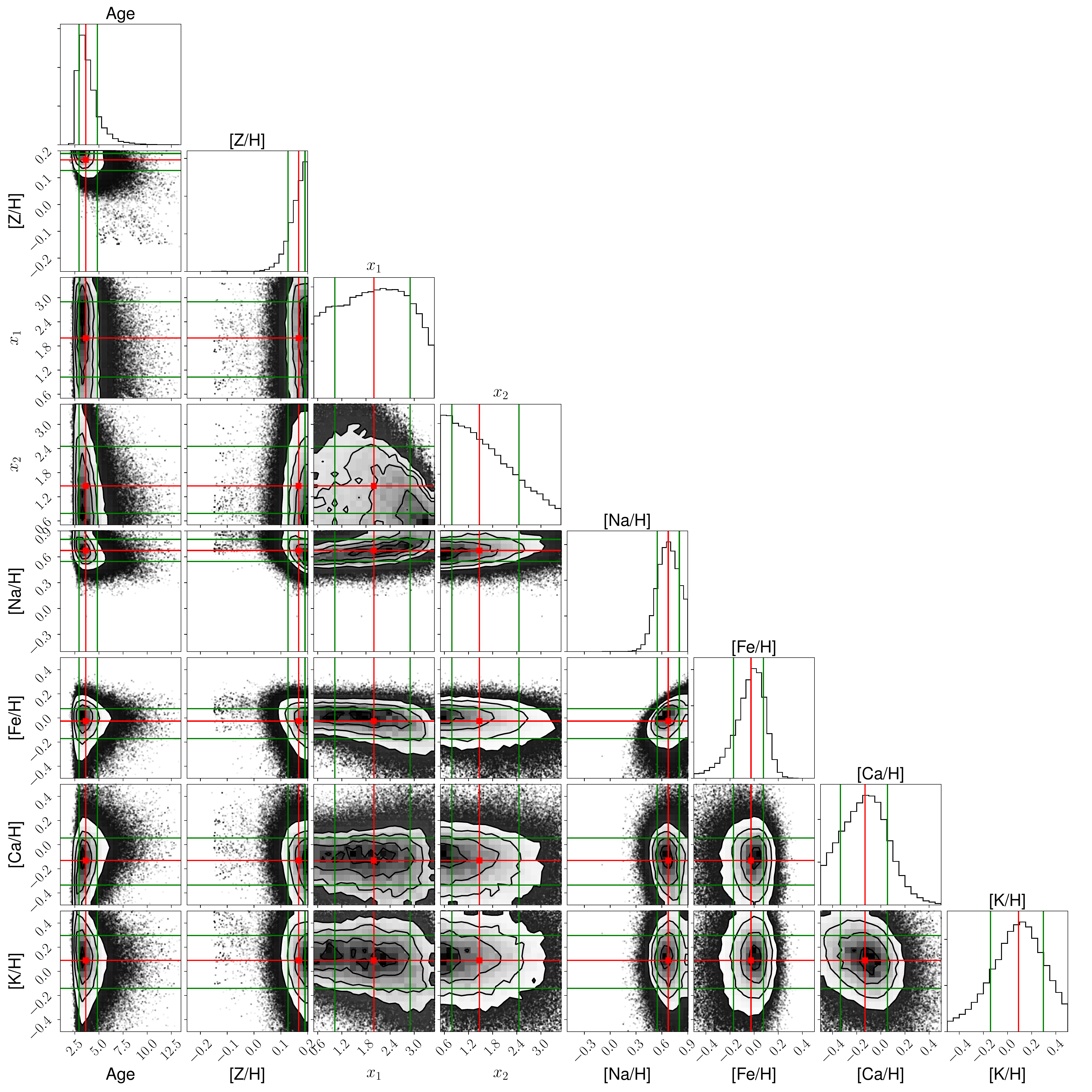}
\caption{`Corner plot' of the MCMC output using the extended set of model parameters derived with the observed M85 spectrum. The extended set of parameters include the Age, [Z/H], the two IMF slopes \imfi{} and \imfii, and four elemental abundance ratios [Na/H], [Fe/H], [Ca/H], [K/H]. The posterior distributions, calculated from the final 1000 samples of each of the 512 MCMC walkers, are shown in the diagonal plots along with the 16$^{th}$, 84$^{th}$ (green lines), and 50$^{th}$ (red lines) percentiles. Below the posterior distributions are co-variance plots (contours) for each pair of parameters. The primary IMF constraint can be seen in the \imfi{}--\imfii{} co-variance plot.\label{fig:MCMCM85}}
\end{figure*}

\begin{figure*}
\epsscale{1.1}
\plotone{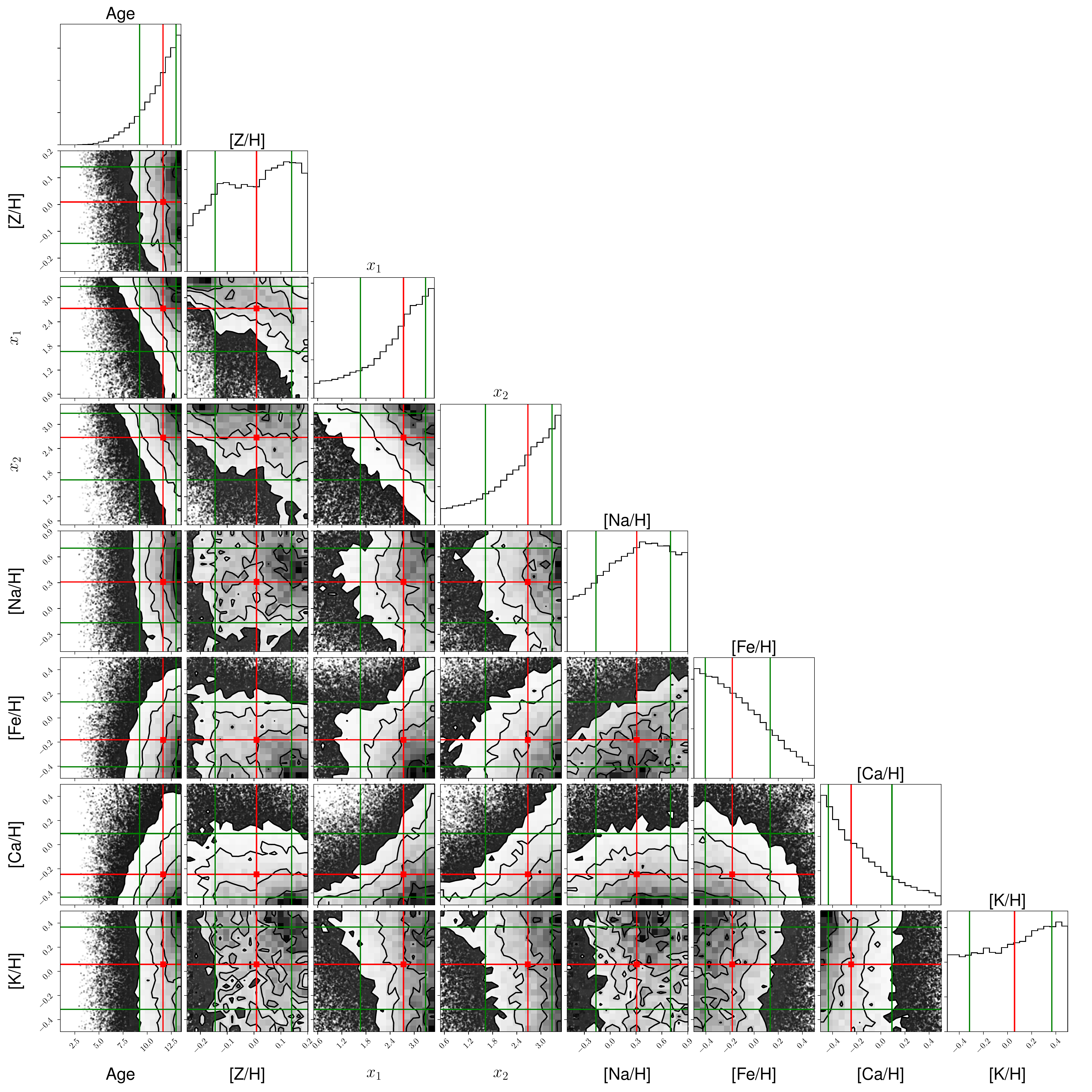}
\caption{`Corner plot' of the MCMC output using the extended set of model parameters derived with the observed M87 spectrum. The layout of this figure is identical to Figure \ref{fig:MCMCM85}.\label{fig:MCMCM87}}
\end{figure*}  

\subsection{Results: MCMC Spectral Fitting}\label{sec:results}
Table \ref{tab:results} presents the `best-fit' model parameters for M85 and M87, as constrained by the final 1000 likelihood samples of the MCMC walkers, including the stellar population ages, metallicities, IMF slopes, abundance ratios, and `IMF-mismatch' parameters ($\alpha_{K}$).
$\alpha_{K}$ is a common measure that relates the constrained (M/L) in the K-band to the (M/L) measured from the models assuming a MW--like IMF.
This parameter is described in more detail in \S\ref{sec:MLR}.
The `best-fit' values are defined as the median (50$^{th}$ percentile level) of the posterior distributions, and the 16$^{th}$ and 84$^{th}$ percentile levels are given as the uncertainties in the median.
Table \ref{tab:results} provides the best-fit results for both the extended and base set of model parameters, and demonstrates that the exclusion of the individual abundance ratios has a significant effect on the measured IMF slopes in M85.
This discrepancy is further discussed in \S\ref{sec:NaM85}.

Figures \ref{fig:MCMCM85} and \ref{fig:MCMCM87} show the posterior distribution plots (diagonals) and the co-variance plots (contours) for the extended set of model parameters, derived from the observed M85 and M87 spectral features, respectively.
The two IMF slopes, \imfi{} and \imfii, have very broad distributions in both figures, indicating that it is difficult to constrain the exact power-law shape of the IMF with our data.
These broad distributions may also be due to the high-degree of co-variance between the individual IMF slopes which results in an unusual best-fit IMF functional form in M85, \imfi{} = 1.98$_{-0.97}^{+0.92}$ and \imfii{} = 1.47$_{-0.69}^{+0.98}$, where the IMF slope is shallower in the range of $0.5\leq M/M_{\sun} <1.0$ than below 0.5 $M/M_{\sun}$.
The best-fit IMF slopes in M87, \imfi{} = 2.74$_{-1.04}^{+0.55}$ and \imfii{} = 2.70$_{-1.01}^{+0.59}$, are consistent with a super--Salpeter IMF.
Despite the broad distributions of the IMF slopes, we find contrasting IMFs in M85 and M87 with the former favouring a more MW-like IMF and the latter more bottom-heavy IMFs.

We measure best-fit ages for M85 and M87 of 3.65$_{-0.72}^{+1.25}$ Gyr and 11.59$_{-2.39}^{+1.40}$ Gyr, respectively. 
The measured ages agree with previous measurements of 3.84\plmn0.78 Gyr and 17.7\plmn2.04 Gyr for M85 and M87 in \citet{ATLAS3D30}.
\citet{ATLAS3D30} argue that their unphysical age measurement for M87 is consistent with the fiducial age of the universe, i.e. $\geq$13.798\plmn0.037 Gyr; \citep{PlanckAde}, when measurement and systematic uncertainties are taken into account.
Furthermore, we note that the stellar population age distribution in Figure \ref{fig:MCMCM87} peaks at the maximum age supplied in the C18 models (13.5 Gyr), which suggests that the stellar age is older than the median age.
Including the \pabeta{} line in the MCMC model analysis has a dramatic effect on the measured best-fit ages.
This effect is discussed in more detail in \S\ref{sec:effectadditional}. 

The best-fit metallicities ([Z/H]) are 0.17$_{-0.04}^{+0.02}$ dex and 0.02$_{-0.16}^{+0.12}$ dex for M85 and M87, respectively.
For M85, the measured metallicity is close to the recent measurement of $[Z/H]~\sim~0.32$ in the galactic core \citep{Ko2018}.
Similar to the measured age of M87, the metallicity posterior distribution for M85 peaks strongly at the C18 model maximum of [Z/H] = 0.2 dex, which suggests that the metallicity is greater than the median value.
The measured metallicity in M87 is consistent with the previous measurement within R$_{e}$/8 from \citet{ATLAS3D30}.

In addition to the IMF slopes, stellar population age, and metallicity, we measure four abundance ratios, [Na/H], [Fe/H], [Ca/H], [K/H], in M85 and M87.
We find evidence for a very high [Na/H] = 0.67$_{-0.13}^{+0.13}$ dex, and a reduced [Ca/H] of --0.13$_{-0.19}^{+0.19}$ dex in M85.
In addition, we measure a [Fe/H] = --0.03$_{-0.15}^{+0.10}$ dex and [K/H] = 0.09$_{-0.22}^{+0.21}$ dex which are consistent with solar metallicity.
We note that the measured [Ca/H] and [Fe/H], are primarily constrained by a single absorption feature.
Their best-fit values are therefore susceptible to degeneracies with other stellar population parameters (e.g. age and the IMF slopes).
The observed M87 spectrum is unable to provide strong constraints on the individual abundance ratios; however, including the abundance ratios does not have a significant effect on the best-fit base parameters in M87.

The addition of the \nai{} 1.27 \micron{} feature does not significantly affect the constrained Na abundance in either galaxy (see \S\ref{sec:NaM85}). 
If the Na abundance is instead constrained with the \nai{} 1.27 \micron{} feature exclusively (i.e. without the \naline{} feature) the best-fit stellar population parameters, including [Na/H], remain unaffected.
The consistency between the Na abundances constrained with either \nai{} feature supports the conclusion that the stellar population in the core of M85 has a significantly enhanced Na abundance.

\subsection{IMF-Mismatch Parameter}\label{sec:MLR}
The individual IMF slopes, \imfi{} and \imfii, show a high degree of correlation regardless of the set of features included in our spectral model fitting.
This can be seen in Figures \ref{fig:MCMCM85} and \ref{fig:MCMCM87} by the consistently broad distributions for the IMF slopes.
To better quantify the IMFs of M85 and M87, we calculate the IMF-mismatch parameter ($\alpha_{K}$) defined as,

\begin{equation}
    \alpha_{K} = ({\rm M/L})_{K}/({\rm M/L})_{K,MW}
\end{equation}

\noindent where (M/L)$_{K}$ is a mass-to-light ratio measured in the K-band, i.e. 2.03--2.37 \micron, and (M/L)$_{K,MW}$ is the K-band mass-to-light ratio assuming an underlying MW-like IMF.
Note that a Kroupa IMF is adopted for the MW-like IMF here.
The (M/L) values are corrected for the remaining stellar mass at the best-fit age, including remnants, following the MIST isochrones \citep{Choi2016}.

The IMF-mismatch parameter is a common measure used to describe the `bottom-heaviness' of IMFs derived from integrated stellar populations \citep[e.g.][]{LaB16, vDC16, Sarzi17}.
$\alpha_{K}$ = 1.0 is representative of a MW-like IMF, while $\alpha_{K} \sim$ 1.8 is equivalent to a Salpeter IMF due to the presence of a higher percentage of low-mass stars that do not contribute significantly to the measured luminosity.
For IMFs steeper than Salpeter, $\alpha_{K}$ increases rapidly, rising to $\alpha_{K} \sim$ 4.0 for an X = 3.0 `bottom-heavy' IMF.

\begin{figure}
\epsscale{1.2}
\plotone{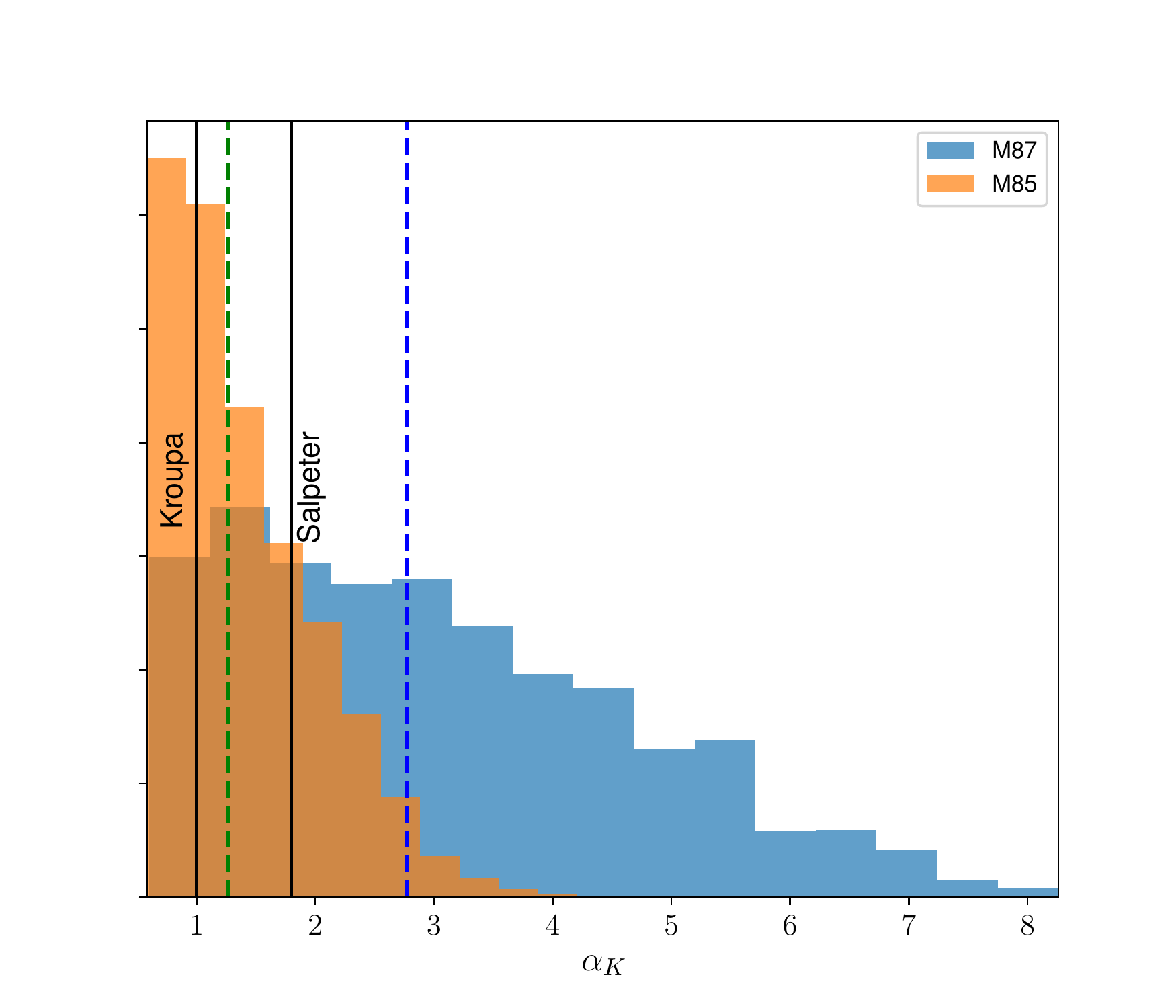}
\caption{IMF-mismatch parameter ($\alpha_{K}$) distributions derived with the best fit model parameters for M85 (orange) and M87 (blue) from Table \ref{tab:results}. The green and blue dashed lines mark the medians of the M85 and M87 distributions, respectively. The solid vertical black lines mark the $\alpha_{K}$ for Kroupa (MW-like) and Salpeter IMFs. The best-fit $\alpha_{K}$ for each galaxy are clearly contrasting with M85 favouring a Kroupa IMF and and M87 a super-Salpeter IMF.\label{fig:MLR}}
\end{figure}

Figure \ref{fig:MLR} shows distributions of $\alpha_{K}$ for both galaxies calculated from models constructed with the best-fit ages and metallicities (Table \ref{tab:results}) and the posterior distributions of the two IMF slopes.
The medians of the $\alpha_{K}$ distributions for M85 and M87 are marked with green and blue dashed lines, respectively.
The median $\alpha_{K}$ values are presented in Tables \ref{tab:results} for each set of stellar population parameters.
We find median $\alpha_{K}$ of 1.26$_{-0.46}^{+0.81}$ and 2.77$_{-1.49}^{+2.09}$ with the best-fit stellar population parameters in Table \ref{tab:results} for M85 and M87, respectively, indicating that the underlying IMFs are likely similar to a Kroupa and super-Salpeter IMF in those galaxies.

\section{DISCUSSION}\label{sec:discussion}

\subsection{Effect of the Stellar Age on the Measured Parameters}\label{sec:effectadditional}

In \S\ref{sec:results}, we noted that the addition of the \pabeta{} feature in the MCMC analysis had a dramatic effect on the constrained stellar ages.
Here we discuss some effects resulting from constraining the model parameters without the \pabeta{} and \newnaline{} features described in \S\ref{sec:additionalfeatures}.
Removing the \pabeta{} feature from the MCMC simulation of the M85 spectra significantly increased the median age from 3.65$_{-0.72}^{+1.25}$ to 7.88$_{-2.90}^{+3.46}$ Gyr.
In contrast to the former, a stellar age of 7.88$_{-2.90}^{+3.46}$ Gyr is inconsistent with recent measurements of the stellar age in the core of M85 from \citet{ATLAS3D30} and \citet{Ko2018}. 
For M87, the median stellar age increased from 10.23$_{-3.44}^{+2.36}$ to 11.59$_{-2.39}^{+1.40}$ Gyr.
This small adjustment was likely a consequence of the stellar age posterior distribution peaking at the maximum 13.5 Gyr regardless of the inclusion of the \pabeta{} feature.

Another important effect of excluding the \pabeta{} and \newnaline{} features from the MCMC simulations was a significant decrease in the best-fit metallicity for M85. 
The decrease in the best-fit [Z/H] in M85, from [Z/H] = 0.17$_{-0.04}^{+0.02}$ in Table \ref{tab:results} to [Z/H] = 0.08$_{-0.12}^{+0.07}$ without the additional features, was likely a consequence of the formerly more tightly constrained, younger stellar population age.
Younger stellar populations are dominated by hotter stars which have intrinsically weaker absorption lines, thereby requiring a higher metallicity to account for the same feature strengths.
As mentioned in \S\ref{sec:results}, the best-fit [Z/H] measured with the additional features was closer to the recent measurement of [Z/H]~$\sim~0.32$ dex in the nucleus of M85 \citep{Ko2018}. 

We find that the best-fit [Z/H] = 0.06$_{-0.16}^{+0.10}$ derived for M87 without the two additional features was increased by approximately 0.04 dex relative to the [Z/H] in Table \ref{tab:results}.
This slight increase was likely also a consequence of the tighter constraints on the stellar population age = 11.59$_{-2.39}^{+1.40}$ Gyr in M87, as an older stellar population reduces the best-fit metallicity given constant index strengths.

Excluding the \pabeta{} and \newnaline{} features from the MCMC simulations did not result in a significant change in the best-fit IMF slopes, which was expected due to the low IMF sensitivity of these features outlined in \S\ref{sec:additionalfeatures}.

\subsection{Impact of elemental abundances on the IMF}\label{sec:NaM85}
In Table \ref{tab:results}, we presented the best-fit stellar population parameters for M85 and M87 when including (extended set) and excluding (base set) the elemental abundances from the MCMC simulations. 
There was a significant discrepancy between the best-fit IMF slopes for M85 when constrained by the base and extended sets of model parameters.
The former is best-fit by a very bottom-heavy IMF while the latter is best-fit by a more MW-like IMF. 
We investigated two possible origins for this inconsistency: the exclusion of individual abundance ratios, i.e., [Fe/H], [Na/H], [Ca/H], [K/H], from the MCMC model parameters, and the removal of individual features (Table \ref{tab:lines}) from the fitting procedure.
When the individual elemental abundances are excluded, their values are fixed to the metallicity ([Z/H]), which does not account for the large deviations of certain elemental abundances such as the high best-fit [Na/H] in M85.
Fixing the abundance ratios to [Z/H] may result in the inconsistency between the IMF slopes when derived with or without the elemental abundance ratios. 
Further MCMC simulations which excluded each of the four line abundance ratios from the model parameters and/or excluded individual IMF-sensitive features from the fitting procedure were therefore conducted to investigate this inconsistency. 

Excluding [Ca/H] and [K/H] from the model parameters did not result in any meaningful differences in the resulting best-fit IMF slopes.
In addition, excluding [Fe/H] resulted in only slightly shallower best-fit IMF slopes in M85.
Individually excluding the FeH, \cai{}, \kai{}, and \ali{} features from the fitting procedure also had no effect on the best-fit IMF slopes.

The best-fit IMF slopes for M85 were significantly steepened when [Na/H] was excluded from the model parameters (i.e. fixing [Na/H] to [Z/H]).
In this case, the best-fit [Fe/H] was dramatically reduced in order to fit the strong \naline{} feature, as this feature strengthens for lower Fe abundances (see \S\ref{sec:sodium}).
If [Fe/H] was excluded in addition to [Na/H] the resulting IMF becomes bottom-heavy, similar to the IMF derived with the base set of model parameters for M85 in Table \ref{tab:results}. 
Furthermore, excluding both [Na/H] and the \naline{} feature from the fitting procedure did not impact the derived IMF slopes.
In summary, we found that including [Na/H] in our model parameters was critical to the interpretation of the strong \naline{} feature in the observed M85 spectrum.

The best-fit IMF slopes for M87 did not exhibit the same sensitivity to the Na abundance and the \naline{} feature.
This was likely a result of both the lower overall S/N in the M87 spectrum and the weaker influence of abundance variations on the spectra of high $\sigma_{v}$ galaxies.

We note here that the strength of the \naline{} feature is known to be difficult to accurately predict with current stellar population models. 
According to \citet{Smith2015}, the strength of this feature was often under-fit despite a strong overall fit to the rest of their observed spectra.
We obtained consistent results, however, when [Na/H] was constrained by either or both of the \naline{} and \newnaline{} features (see \S\ref{sec:results}).
This consistency, along with the assertion in \citet{Smith2015} that the \newnaline{} feature is only visible at high [Na/H], increased our confidence in the high [Na/H] reported in Table \ref{tab:results} for M85.

\subsection{M87 AGN Correction}\label{sec:AGNCorrection}
M87 is known to have a strong, central AGN that contaminates the observed spectral continuum (see \S\ref{sec:reduction}).
The \kai{} 1.25 \micron{} line was excluded from our analysis of the stellar population properties in M87 due to a [Fe II] emission line from this AGN.
We considered two possible effects of accounting for the M87 AGN on the best-fit IMF slopes: excluding the \kai{} 1.25 \micron{} feature in our analysis of M85, and correcting the M87 spectrum for an estimated AGN continuum emission. 
For the former, we repeated the MCMC analysis of the M85 spectrum while excluding the \kai{} 1.25 \micron{} feature and found that the best-fit IMF slopes are not significantly affected. 

To measure the effect of removing the AGN continuum, we repeated our analysis of M87 after subtracting a percentage of the linear continuum defined for each spectral feature in Table \ref{tab:lines}.
We adopted the estimation from ASL17 that the AGN in M87 contributes approximately 15\% of the observed continuum level within the central 3\arcsec{}.
As expected, this correction resulted in a steepening of the best-fit IMF slopes in M87 from a Salpeter-like (X $\sim$ 2.3) to a bottom-heavy (X $\sim$ 3.0) IMF.
This did not affect our overall conclusion that the best-fit IMF in M87 is consistent with a Salpeter or steeper IMF.

\subsection{Comparison to Previous IMF Results}\label{sec:comparison}
The results presented in \S\ref{sec:MCMCResults} represent the first measurement of contrasting IMFs in extragalactic stellar populations, with this set of NIR, IMF-sensitive absorption features.
In general, the best-fit IMFs for M85 and M87 presented in this paper are consistent with previous measurements of the IMFs in those galaxies.
This is important as it shows that constraints on the IMF are robust against the specific set of IMF-sensitive features or spectral bandpass. 
The work in this paper also provides a method for investigating the IMF of stellar populations with NIR spectroscopy.
In this section, we compare our results to previous measurements of stellar population parameters in M85 and M87, e.g. vDC12 and \citet{Capp2013}.

The IMF slopes obtained for both of the galaxies are generally consistent with the IMFs derived from long--slit spectroscopy in the visible bands in CvD12b and galactic kinematics in \citet{Capp2013}.
CvD12b measured a `bottom-light' IMF in M85 with $\alpha_{K}$ = 0.63, which is shallower than the MW-like $\alpha_{K}$ = 1.26$_{-0.46}^{+0.81}$ reported in Table \ref{tab:results}.
We note, however, that the $\alpha_{K}$ distribution for M85 in Figure \ref{fig:MLR} peaks strongly at the lowest $\alpha_{K}$, suggesting that the $\alpha_{K}$ for M85 is lower than the median.
In M87, CvD12b found a super-Salpeter IMF with $\alpha_{K}$ = 1.90 that is similar to our median $\alpha_{K}$ = 2.77$_{-1.49}^{+2.09}$.
\citet{Capp2013} measured the IMFs in M85 and M87 to be slightly steeper than a Kroupa IMF and slightly shallower than a Salpeter IMF, respectively, which is qualitatively consistent with our measurements.

Our measurement of a super-Salpeter IMF in the center of M87 is similar to the results of \citet{Sarzi17} and \citet{Oldham18}.
Using spectroscopy in the visible bands, \citet{Sarzi17} measured an IMF slope of approximately 2.9 in the core of M87 which is consistent with both IMF slopes for M87 in Table \ref{tab:results} of $\sim$2.7.
\citet{Oldham18} inferred a single, power-law slope below 1.0 M$_{\sun}$ of approximately 2.5 in the core of M87 with a sophisticated dynamical model constructed from observations of M87 satellites.

A novel result of this paper is the measurement of four abundance ratios in M85, [Na/H], [Fe/H], [Ca/H], and [K/H], of which only [Fe/H] has prior reported measurements.
vDC12 measured [Fe/H] = -0.02 dex in M85 which is consistent with our [Fe/H] measurement of $-0.03_{-0.15}^{+0.10}$ dex in Table \ref{tab:results}.

We measured an exceptionally enhanced [Na/H] = 0.67$_{-0.13}^{+0.13}$ dex in M85, which is typical of the cores of old, massive ETGs \citep{Spiniello2012}.
M85, however, is known to have a young, counter-rotating, kinematically decoupled core within the central 1\arcsec{} that may have formed from a recent, wet galaxy merger \citep{McDermid2004, Terlevich2002}.
This burst of star formation in the core of M85 may be the cause of the enhanced Na abundance.
Na is injected into the interstellar medium (ISM) by both the stellar winds of massive stars and by \ion{Type}{2} supernovae, the latter has a strong, metallicity-dependent yield.
Assuming our best-fit [Z/H] = 0.09 dex is accurate, the high Na abundance in M85 may be a result of this metallicity-dependent Na yield \citep{Kobayashi2006}.
The Na enriched gas ejected into the ISM from recently formed massive stars may be accreted onto existing or still forming stars, due to the high stellar density in the core of M85 \citep{McConnell2016}.

Furthermore, we measured a strong \ali{} 1.31 \micron{} index in M85.
Al and Na are produced in a similar fashion during the carbon burning phase of massive stars \citep{Lecureur07}.
Al also exhibits the same strong metallicity-dependent yield in Type II supernovae as Na.
It is therefore possible that both the strong [Na/H] and the high \ali{} index strength are a consequence of the recent star formation in M85.

Na is also produced as a product of hot bottom burning in intermediate mass (3--8 M$_{\sun}$) AGB stars \citep{Vaughan2018}. 
There is evidence that the Na yield from this process also increases with metallicity \citep{Ventura2013}. 
At $\sim$3.5 Gyr, these stars have already enriched the ISM and may have further contributed to the high [Na/H] measured in M85.

Due to the lower S/N of the M87 spectra, we are not able to place strong constraints on the individual abundance ratios; however, we note that the best-fit [Na/H] = 0.34$_{-0.50}^{+0.37}$ dex for M87 (Table \ref{tab:results}) is marginally consistent with recent measurements of a high [Na/H] = $\sim 0.7$ dex in the central regions of the galaxy \citep{Sarzi17}.

A key question is, if the IMF is varying in ETGs as described by the stellar population models, what is the astrophysical driver for this variation?
The main parameters proposed as the origin for the observed IMF variation in ETGs are $\sigma_{v}$ and metallicity (see \S\ref{sec:introduction}).
The IMFs measured in this paper give further support to the observed trend of increasingly bottom-heavy IMFs in ETGs with higher central $\sigma_{v}$ (CvD12b).
Recently, \citet{Parikh2018} measured correlations of varying degrees between the low mass ($\leq$ 0.5 \msun) IMF slope and galactic properties such as $\sigma_{v}$, [Z/H], [Na/Fe], galactic radius, and stellar age in a sample of $\sim$400 ETGs.
Our best-fit \imfi{} slope for M85 and M87 is consistent with their IMF--$\sigma_{v}$ correlation.
The \citet{Parikh2018} sample, however, only includes galaxies with $\sigma_{v} < 200$ \kms, so the comparison to our measured IMF slopes in M87 is only qualitative.
\citet{Parikh2018} also found that the IMF in galaxies with high [Z/H] and [Na/Fe] are consistent with a Salpeter IMF, which is only marginally consistent with the \imfi{} slope measured for M85 in this paper.
In particular, they measured a very tight correlation between the IMF slope and [Z/H], which is inconsistent with our IMF slope in M85.
A possible explanation for this inconsistency is the lack of galaxies younger than 5 Gyr in the sample studied by \citet{Parikh2018}.

The measurements of bottom-heavy IMFs in the core of massive galaxies, as measured for M87 in this paper, and more MW-like IMFs in less-massive galaxies like M85, implies that the conditions during the formation of these stellar systems were fundamentally different.
The cores of brightest cluster galaxies, such as M87, were likely formed in massive dark matter potential wells and experienced an exceptionally dense star formation environment \citep{Oldham18}.
The stellar populations in the cores of these galaxies would therefore be extremely old, as measured for M87 in Table \ref{tab:results}.
In contrast, M85 is a much less massive galaxy with a younger stellar population, comparable to those in the Milky Way.
The stars in the core of M85 likely formed in a significantly different environment than those in the core of M87. 
\citet{Chabrier2014} examined the physical basis for this type of IMF variation and determined that the compressive turbulent motions found in extreme star formation environments can shift the characteristic mass of the IMF to a lower mass, resulting in a bottom-heavy IMF.
This provides a consistent picture for the origin of the observed IMF variation for M85 and M87 in this paper, as a function of their contrasting formation histories.

\section{SUMMARY AND FUTURE WORK}\label{sec:conclusions}
In this paper, we have presented a study of the IMFs for two ETGs, M85 and M87, with highly contrasting central velocity dispersions and [$\alpha$/Fe] using a set of seven NIR, gravity-sensitive absorption features.
This set of features have been relatively unexplored for the purpose of measuring variations in the IMF from the integrated light of extragalactic stellar populations. 
To that end, we compared the observed spectral regions for these features for both galaxies to stellar population models described in C18.

Our key conclusions are as follows: 

\begin{itemize}
    \item Our measured feature indices and median $\alpha_{K}$ for M85 and M87 are consistent with those found in previous studies using both spectroscopic and kinematic techniques \citep[e.g.,][vDC12, ASL17]{Capp2013, Sarzi17}.
    \item The EW index strengths of the seven IMF-sensitive features give inconsistent constraints for the IMF slopes in both galaxies relative to fiducial models defined with previously measured stellar population ages and metallicities. These inconsistencies between the predicted IMF slopes and the measured index strengths can be reconciled by assuming particular elemental abundance ratios, thereby underlining the necessity of considering abundance variations when investigating the IMF.
    \item The best-fit IMF slopes in M85, derived with the MCMC model fitting, are consistent with a MW-like IMF. However, the best-fit \imfi{} slope is steeper than the \imfii{} slope which describes a unusual IMF functional form. This unusual IMF is likely a consequence of the high co-variance between the individual IMF slopes. The median IMF-mismatch parameter, $\alpha_{K} = 1.26$, allows for a more definitive interpretation of a MW-like IMF in M85. 
    \item The best-fit IMF slopes in M87 are consistent with a super-Salpeter IMF. The median $\alpha_{K}$ = 2.77 supports our conclusion that the IMF in M87 is likely between a Salpeter and a bottom-heavy IMF. The MCMC simulations were unable to constrain the individual abundance ratios in M87 due to the lower S/N of the M87 spectrum than the M85 spectrum. If the M87 spectral continuum is corrected for an estimated 15\% additional continuum from the central AGN, the best-fit IMF slopes steepen to a bottom-heavy IMF. 
    \item We find a significantly enhanced Na abundance ([Na/H]~$\sim$~0.65 dex) in M85. The measured abundance ratio is consistent when constrained with either of the considered \nai{} features. This high [Na/H] may be a consequence of both the high metallicity and the recent burst of star formation in the core of the M85. We conclude that, due to the high Na abundance, including [Na/H] in the model parameters is critical to the interpretation of the IMF slopes in M85.
\end{itemize}

This work adds to the growing body of evidence that the IMFs of ETGs vary as a function of fundamental galactic properties (e.g. $\sigma_{v}$, [$\alpha$/Fe], [Z/H]) while also illustrating the viability of using NIR IMF-sensitive features as possible tools for investigating IMF variation.
We are currently conducting a survey of nearby ETGs and the bulges of spiral galaxies using the recently commissioned Wide Integral Field Infrared Spectrograph \citep[WIFIS:][]{WIFIS1,WIFIS2}. 
Similarly to the NIFS spectrograph used in this work, WIFIS operates in the zJ-band and will be able to perform a spatially resolved investigation of the IMF slopes out to large radii using this set of IMF-sensitive features. 
This survey will serve as a NIR companion to integral field IMF variation studies in the visible bands such as the MANGA \citep{Bundy2014}, CALIFA \citep{Sanchez2012}, and ATLAS3D \citep{Cappellari2011} surveys. 
We anticipate that our analysis of the observed galaxies in the WIFIS extragalactic survey will significantly contribute to our broader understanding of IMF variation in nearby extragalactic objects.

\acknowledgements
We thank C. Conroy and his group for the use of their recent stellar population models in this paper. DSM was supported in part by a Leading Edge Fund from the Canadian Foundation for Innovation (project No. 30951). Both DSM and SS were supported by a Discovery Grant from the Natural Sciences and Engineering Research Council of Canada (NSERC). We thank the anonymous referee for their helpful comments which improved the paper.

This paper is based on observations obtained at the Gemini Observatory, which is operated by the Association of Universities for Research in Astronomy, Inc., under a cooperative agreement with the NSF on behalf of the Gemini partnership: the National Science Foundation (United States), the National Research Council (Canada), CONICYT (Chile), Ministerio de Ciencia, Tecnolog\'{i}a e Innovaci\'{o}n Productiva (Argentina), and Minist\'{e}rio da Ci\^{e}ncia, Tecnologia e Inova\c{c}\~{a}o (Brazil).

The analysis in this publication made extensive use of the \textsc{python} modules NumPy \citep{Numpy}, AstroPy \citep{Astropy}, SciPy \citep{Scipy}, and Matplotlib \citep{matplotlib}.

\end{document}